\documentclass{aa}

\usepackage{txfonts}

\usepackage{subcaption}
\usepackage{cuted}
\usepackage{capt-of}
\usepackage{natbib}
\usepackage{hyperref}
\usepackage{graphicx}
\usepackage{amsmath}
\usepackage{amssymb}
\usepackage{bm} 
\usepackage{amsfonts}
\usepackage{amstext}
\usepackage{diagbox}
\usepackage{makecell}
\usepackage{placeins}
\usepackage{url}
\usepackage{longtable}
\usepackage{booktabs}
\usepackage{array}
\usepackage{multirow}
\usepackage{xcolor}
\usepackage{ulem}
\usepackage{mathrsfs}
\usepackage{tabularx}
\usepackage{subcaption}
\usepackage{afterpage}

\usepackage{longtable}
\usepackage{placeins}
\renewcommand{\d}{\mathrm{d}}
\newcommand{\e}{\mathrm{e}}

\newcommand{\bea}{\begin{eqnarray}}
\newcommand{\eea}{\end{eqnarray}}
\newcommand{\be}{\begin{equation}}
\newcommand{\ee}{\end{equation}}

\renewcommand{\exp}{\mathrm{exp}}

\newenvironment{revised}{\par}{\par}

\nolinenumbers\let\linenumbers\relax
\begin{document}

\title{The mass density slope of early-type galaxies: newest results
       from the redshift distribution of strong gravitational lenses}

\author{Xinyue Jiang\inst{1,2,4}
   \and Shuo Cao\inst{1,2}\corrauth{caoshuo@bnu.edu.cn}
   \and Wuzheng Guo\inst{1,2}
   \and Chenfa Zheng\inst{1,2}
   \and Chengsheng Mu\inst{1,2}
   \and Shuaibo Geng\inst{3,4}
   \and Marek Biesiada\inst{4}
   \and Yalong Nan\inst{1,2}
   }

\institute{School of Physics and Astronomy, Beijing Normal University,
           Beijing 100875, China;
   \and    Institute for Frontiers in Astronomy and Astrophysics,
           Beijing Normal University, Beijing 102206, China;
   \and    School of Physics and Electronics, Hunan Normal University,
           Changsha 410081, China;
   \and    National Centre for Nuclear Research, Pasteura 7,
           02-093 Warsaw, Poland
   }

\date{Received XXX; accepted XXX}

\abstract
{}
{The mass distribution of early-type galaxies (ETGs) encodes key information about baryonic processes and dark-matter assembly. In this paper, we propose a new framework to investigate the mass-density profiles of ETGs and their evolution with redshift, based on the redshift distribution of strong gravitational lenses.}
{We compile a large set of 200 galaxy-scale strong lensing systems with complete information, including lens and source redshifts, Einstein radii, and spectroscopic velocity dispersions. To account for the heterogeneous construction of the compiled sample, we model a lens-redshift-dependent selection function for each survey category. By combining the lens-redshift distribution with velocity dispersion function (VDF) and spherical Jeans-based mapping, we obtain population-level constraints on the total mass-density slope $\gamma$ in three lens-redshift intervals, marginalizing over the luminosity-density slope $\delta$ and the stellar orbital anisotropy $\beta$.}
{We find that the total mass density of ETGs remains close to isothermal at low and intermediate redshifts ($\gamma=2.00^{+0.07}_{-0.15}$ at $0<z_l<0.5$ and $\gamma=1.96^{+0.16}_{-0.26}$ at $0.5<z_l<0.6$), with a shallower effective slope at higher redshift ($\gamma=1.81^{+0.17}_{-0.21}$ at $0.6<z_l\lesssim1.0$). The marginalized posteriors of $\delta$ and $\beta$ remain broad, consistent with $\delta \simeq 2$ and with isotropic orbits in all three redshift intervals. We also perform several diagnostic tests to confirm the stability of the inferred density slopes. These results are consistent with the redshift trend predicted by Horizon-AGN simulations, in which active galactic nucleus (AGN) feedback, together with stellar assembly and merging, contributes to the structural evolution of ETGs. Our findings demonstrate that the population-level total mass-density slope of ETGs can be constrained from the redshift distribution of existing strong-lens samples. Applied to the much larger lens samples expected from future missions such as the Chinese Space Station Telescope (CSST), this method will enable substantially tighter tests of density-slope evolution and its connection to galaxy formation processes.}
{}

\keywords{gravitational lensing: strong --
          galaxies: evolution --
          galaxies: structure 
         }

\titlerunning{Density slope of ETGs from lens redshift test}
\authorrunning{Jiang et al.}

\maketitle

\section{Introduction} \label{sec:intro}

The mass distribution of galaxies has been the focus of many observational studies, as it provides a direct probe of both galaxy-formation physics and the connection between luminous and dark matter \citep{2015MNRAS.451.1247S,2020NatRP...2...42V,2024ARNPS..74..173P}. On galactic scales, the inner density profile reflects the coupled evolution of baryons and dark matter, offering key insights into star-formation efficiency, feedback, halo response, and the assembly history of galaxies \citep{1991ApJ...379...52W,2014A&A...562A..30S}. These processes redistribute matter within galaxies and leave observable imprints on velocity dispersion, stellar mass-to-light ratio, and dynamical structure. Radial mass profiles are also essential for connecting galaxy-scale measurements to halo modeling and structure formation in the $\Lambda$CDM framework \citep{2017MNRAS.467.2019R,2017MNRAS.466.1648K,2024MNRAS.528.1655J}, and they provide useful tests of dark-matter physics on sub-galactic to halo scales \citep{2017MNRAS.464.4520B,2018PhR...730....1T,2023MNRAS.524.1515M}. Among different galaxy populations, early-type galaxies (ETGs) are particularly well suited for such studies because their quiescent stellar populations, high central densities, and relatively regular morphologies make them favorable targets for both dynamical and strong lensing analyses \citep{2006ApJ...649..599K,2009ApJ...705.1099A}. The total density slope of ETGs therefore provides a compact and physically informative quantity with which to study the interplay between baryonic assembly and dark-matter structure across cosmic time. Massive ETGs are often found to have total mass-density profiles that are close to isothermal over the radial range probed by stellar dynamics and strong lensing \citep{2015ApJ...804L..21C,2016ARA&A..54..597C}. This near-isothermal structure is commonly interpreted as the result of the coupled contribution of a centrally concentrated stellar component and a more extended dark-matter halo, shaped by baryonic assembly, feedback, quenching, and merging processes \citep{2010ApJ...725.2312O,2007ApJ...658..710N,2008MNRAS.384....2G}. Several theoretical and empirical profiles have been used to describe galaxy mass distributions. The Navarro-Frenk-White (NFW) profile, motivated by collisionless $\Lambda$CDM simulations, provides a useful description of dark-matter halos, but the inner regions of massive galaxies are strongly affected by baryonic physics \citep{2014MNRAS.437..415D,2021MNRAS.501.5679C}. For lensing and dynamical studies of ETGs, the total mass distribution is therefore often described by an effective power-law profile, $\rho_{\rm tot}(r)\propto r^{-\gamma}$, where $\gamma=2$ corresponds to the isothermal case \citep{2016MNRAS.456..739X,2018MNRAS.480..431L,2019MNRAS.487.5143O}. This compact parametrization captures the combined stellar and dark-matter distribution over the radial range most relevant for strong lensing and provides a useful way to quantify possible structural evolution.

Several complementary observational methods have been used to measure ETG mass profiles and to test whether their effective total density slopes evolve with redshift. Stellar-dynamical analyses infer the three-dimensional mass distribution from stellar motions or other tracers, typically through Jeans or orbit-based modeling and with assumptions about the tracer distribution and orbital anisotropy \citep[e.g.][]{2017MNRAS.467.1397P,2021MNRAS.506.3691D}. These studies provide direct dynamical information, although high-quality spectroscopy, especially spatially resolved spectroscopy at high redshift, remains observationally demanding \citep{2014ApJ...788...28V,2018ApJ...862..126N}. Strong gravitational lensing provides an independent and complementary constraint through the projected mass enclosed within the Einstein radius, however, lensing-only inferences of density slopes are affected by degeneracies such as the mass-sheet degeneracy and the coupling between density slope and external convergence \citep[e.g.][]{1985ApJ...289L...1F,1992grle.book.....S}. Joint lensing-and-dynamics (L\&D) analyses combine these complementary observables and have provided some of the strongest constraints on individual ETG lenses \citep[e.g.][]{10.1111/j.1365-2966.2011.18842.x,2018MNRAS.480..431L,2006ApJ...649..599K}. Despite the broad agreement that massive ETGs are approximately isothermal, the reported normalization and redshift dependence of $\gamma$ vary among samples and methods. Some lensing-only analyses have reported evidence for a positive trend of $\gamma$ with redshift \citep{2023MNRAS.521.6005E}, whereas several joint L\&D analyses have found a decreasing total density slope with redshift \citep{2011ApJ...727...96R,2012ApJ...757...82B,2013ApJ...777...98S,2025A&A...694A.196G,2025MNRAS.541....1S}. Dynamics-only studies of nearby ETGs have often reported little or no redshift evolution in the inferred total slope \citep[e.g.][]{2017MNRAS.467.1397P,2021MNRAS.506.3691D,2023MNRAS.522.3602D}. However, the recent analysis of \citet{Mozumdar2026}, based on deep Multi Unit Spectroscopic Explorer (MUSE-DEEP) spectroscopy and the Mapping Nearby Galaxies at Apache Point Observatory (MaNGA) survey, reports a mild but $\sim5\sigma$-significant negative evolution of $\mathrm{d}\gamma/\mathrm{d}z \approx -0.20\pm0.03$, a dynamics-only result that is closer in sign to the negative evolution reported by several lensing-dynamics studies. These differences may reflect genuine departures from a single power-law description \citep{2023MNRAS.521.6005E}, differences in the radial range probed by each technique, or sample-selection effects. Measuring the population-level redshift dependence of $\gamma$ therefore remains important for understanding ETG structural evolution and for assessing the assumptions entering strong lensing statistics.

The rapid growth of strong-lens samples is opening a new regime for such measurements. Spectroscopic searches in the Sloan Digital Sky Survey (SDSS) established well-characterized samples of galaxy-scale lenses \citep{2008ApJ...682..964B}, while wide-field imaging surveys have produced increasingly large catalogues of lens candidates, including those from the Hyper Suprime-Cam (HSC) Subaru Strategic Program and the Ultraviolet Near-Infrared Optical Northern Survey (UNIONS) \citep{2018PASJ...70S..29S,2022A&A...666A...1S,2025A&A...693A.291S,2026ApJ...999....1S}. Recent work has further demonstrated the value of combining imaging-based searches with wide-area spectroscopy, as illustrated by the spectroscopic confirmation of HSC lens candidates with the Dark Energy Spectroscopic Instrument (DESI) \citep{2025SCPMA..6829511S}. The first Euclid Quick Data Release alone yielded 497 galaxy-scale lens candidates over $63\,{\rm deg}^2$ \citep{2026A&A...711A..26E}, illustrating the scale of the samples that will become available from Euclid and other forthcoming wide-field surveys. Forecasts for Euclid, the Vera C. Rubin Observatory Legacy Survey of Space and Time (LSST), Roman, and the Chinese Space Station Telescope (CSST) indicate that these facilities may together deliver lens samples of order $10^5$ systems \citep{2015ApJ...811...20C,2020RNAAS...4..190W,2024MNRAS.533.1960C}. The resulting samples will span different discovery channels, imaging depths, angular resolutions, and spectroscopic follow-up strategies, making the treatment of survey selection an increasingly important component of population-level lensing analyses \citep{2024SSRv..220...23L}.

In this work, we use strong lensing statistics to extend the lensing-dynamics connection to the population level by combining the lens-redshift distribution with the velocity-dispersion function (VDF) of the lens population \citep{2003MNRAS.344.1029D}. Rather than performing detailed dynamical modeling for each individual lens, we apply a statistical strong lensing framework in which ensemble lensing observables are combined with VDF calibrations and a Jeans-based mapping. For our newly compiled sample of 200 galaxy- scale ETG lenses drawn from different discovery channels, we incorporate a redshift-dependent selection function into the lens-redshift likelihood and investigate the mass profiles of ETG lenses and their possible redshift dependence. The paper is organized as follows: Section~\ref{sec:Meth} describes the lensing statistics and the selection function used in our analysis. Section~\ref{sec:Data} introduces the observational strong lensing sample. Section~\ref{sec:results} presents the inferred density slopes and stability tests, based on the redshift distribution of available gravitational lenses. Section~\ref{sec:Conclusion} summarizes the main conclusions and discussion, forecasting the performance of our methodology in future missions such as CSST. Appendix~\ref{app:SB} gives the details of the selection function calculation.

\section{Methodology} 
\label{sec:Meth}

\subsection{Strong lensing optical depth and the lens galaxy model}

Our method utilizes the strong lensing optical depth, which describes the lensing probability density for a background source at $z_s$ being lensed by a foreground galaxy at redshift $z_l$. Since galaxy-scale strong lensing events are rare, the lensing probability can be described in terms of an optical-depth formalism. Following the framework of \citet{2003MNRAS.343..639O,2012A&A...538A..43C,2012ApJ...755...31C}, we write the differential optical-depth kernel evaluated at a given Einstein radius $\theta_{\rm E}$ as
\begin{equation}  
\frac{d\tau}{dz_l}(\theta_{\rm E},z_l|z_s) = n(\theta_{\rm E}, z_l) \cdot (1+z_l)^3 \cdot S_{\rm cr} \cdot \frac{c\,dt}{dz_l} \,,  
\label{eq:dtao_dzl}  
\end{equation}  
where $n(\theta_{\rm E}, z_l)$ is the comoving number density of lensing galaxies characterized by Einstein radius $\theta_{\rm E}$ at lens redshift $z_l$, and the factor $(1 + z_l)^3$ converts it to the corresponding physical number density. The lensing cross section is approximated as
\begin{equation}  
S_{\rm cr} = \pi (\theta_{\rm E} D_l)^2,   
\end{equation} 
where $D_l$ denotes the angular diameter distance from the observer to the lens galaxy. This term represents the characteristic lensing area associated with the observed Einstein radius. The line-of-sight path-length element is
\begin{equation}  
\frac{cdt}{dz_l} = \frac{c}{H(z_l)(1+z_l)} \, ,
\end{equation}  
where $H(z_l)$ denotes the Hubble parameter at redshift $z_l$. Throughout this work we adopt a flat $\Lambda$CDM cosmology with $\Omega_m=0.3$ and $H_0=70\,{\rm km\,s^{-1}\,Mpc^{-1}}$.

Based on the framework proposed by \citet{2006EAS....20..161K}, we adopt a spherical, scale-free model for the lens galaxy, in which the stellar luminosity density $\nu(r)$, total mass density $\rho(r)$, and orbital anisotropy $\beta$ are parameterized as
\begin{equation}
\nu(r)=\nu_{0}r^{-\delta},\quad
\rho(r)=\rho_{0}r^{-\gamma},\quad
\beta(r)=1-\frac{\langle\sigma_\theta^2\rangle}{\langle\sigma_r^2\rangle}.
\end{equation}
Here $\nu_0$ and $\rho_0$ are normalization constants, and $\sigma_r$ and $\sigma_\theta$ are the radial and tangential components of the stellar velocity dispersion. In this parameterization, $\gamma$ describes the total mass-density slope, $\delta$ describes the stellar luminosity-density slope, and $\beta$ is the constant orbital anisotropy parameter. This scale-free form provides a compact phenomenological description that is well suited for statistical lensing applications.

Within this framework, the spherical Jeans equation provides the connection between the mass distribution and the stellar kinematics. Specifically, it links the stellar velocity dispersion $\sigma_{\rm ref}$, defined with respect to a reference radius $R_{\rm ref}$, to the Einstein radius of the lens galaxy, $R_{\rm E}=D_l\theta_{\rm E}$. Depending on the dataset, $R_{\rm ref}$ may correspond to the spectroscopic aperture radius $R_{\rm ap}$, the effective radius $R_e$, or a standardized aperture such as $R_e/8$. Following \citet{2006EAS....20..161K}, the relation can be expressed as
\begin{equation}
\sigma_{\rm ref}^2
=\frac{1}{\pi}\left(\frac{G\,M_{\rm dyn}}{R_{\rm E}}\right)
\frac{1}{f(\gamma,\delta,\beta)}
\left(\frac{R_{\rm ref}}{R_{\rm E}}\right)^{2-\gamma},
\label{eq:th_ap}
\end{equation}
where
\(\displaystyle \xi \equiv \gamma+\delta-2\) and 
\begin{equation}
\begin{aligned}
f(\gamma,\delta,\beta)
&=\Bigg[
\frac{2\sqrt{\pi}\,(\delta-3)}{(\xi-3)(\xi-2\beta)}
\frac{\Gamma(\delta/2)\,\Gamma(\gamma/2)}
{\Gamma[(\delta-1)/2]\,\Gamma[(\gamma-1)/2]}
\\ &\hspace{1.5em}\times
\left(
\frac{\Gamma[(\xi-1)/2]}{\Gamma(\xi/2)}
-\beta\,\frac{\Gamma[(\xi+1)/2]}{\Gamma[(\xi+2)/2]}
\right)
\Bigg]^{-1}.
\end{aligned}
\label{eq:f_g_d_b}
\end{equation}
When assuming $\gamma=\delta$ and $\beta=0$, the commonly used $\gamma$-only approximation can be written as
\begin{equation}  
\begin{aligned}
f(\gamma) & =-\frac{1}{\sqrt{\pi}} \frac{(5-2 \gamma)(1-\gamma)}{3-\gamma} \frac{\Gamma(\gamma-1)}{\Gamma(\gamma-3 / 2)} \\
& \times\left[\frac{\Gamma(\gamma / 2-1 / 2)}{\Gamma(\gamma / 2)}\right]^2 \, .
\end{aligned}  
\label{eq:f_gamma} 
\end{equation} 
For $\gamma=2$ in Eq.~(\ref{eq:f_gamma}), Eq.~(\ref{eq:th_ap}) reduces to the singular isothermal sphere (SIS) limit, with the velocity-dispersion relation $\sigma_{\rm ref}^2(R) = \frac{1}{\pi}\left[GM_{\rm dyn}/R_{\rm E}\right]$. We follow the same dynamical mass--Einstein radius formulation adopted in \citet{2015ApJ...806..185C}. The dynamical mass associated with the Einstein radius, derived from Eq.~(\ref{eq:th_ap}), is expressed as
\begin{equation}
\begin{aligned}
M_{\rm dyn} & = \frac{\pi}{G} \sigma_{\rm ref}^2 R_{\rm E} \left(\frac{R_{\rm E}}{R_{\rm ref}}\right)^{2-\gamma} f(\gamma,\delta,\beta) \\
& = \frac{\pi}{G} \sigma_{\rm ref}^2 D_l \theta_{\rm E} \left(\frac{\theta_{\rm E}}{\theta_{\rm ref}}\right)^{2-\gamma} f(\gamma,\delta,\beta),
\label{eq:Mdyn}
\end{aligned}
\end{equation}
where $\theta_{\rm ref} \equiv R_{\rm ref}/D_l$ denotes the corresponding angular reference radius. The lensing mass projected within the Einstein radius follows from the critical surface density \citep{Meneghetti2021IntroductionTG}
\begin{equation}
M_{\rm lens} = \frac{c^2}{4G} \frac{D_s D_l}{D_{ls}} \theta_{\rm E}^2,
\label{eq:Mlens}
\end{equation}
where $D_s$ and $D_{ls}$ represent the angular diameter distances from the observer to the source and from the lens to the source, respectively. The Einstein radius is a robust lensing observable and is less sensitive than the full mass model to the detailed radial parameterization of the lens potential \citep{1985ApJ...289L...1F,2017A&A...601A..77U,2024A&A...692A..87G,2022MNRAS.517.3275E}. This makes it suitable for a statistical analysis of heterogeneous lens samples.

From the equations above, the reference angular scale $\theta_{\rm ref}$ enters Eq.~(\ref{eq:Mdyn}) through the factor $(R_{\rm ref}/R_{\rm E})^{2-\gamma}$, which accounts for the aperture dependence of the velocity-dispersion normalization. By combining the dynamical mass in Eq.~(\ref{eq:Mdyn}) with the lensing mass inferred from Eq.~(\ref{eq:Mlens}), we obtain an explicit relation between the mass-density slope $\gamma$ and the Einstein radius,
\begin{equation}  
\theta_{\rm E} = \left[4\pi \left(\frac{\sigma_{\rm ref}}{c}\right)^2 \theta_{\rm ref}^{\gamma-2} \frac{D_{ls}}{D_s} f(\gamma,\delta,\beta) \right]^{\frac{1}{\gamma-1}}. 
\label{eq:thE_sigma}  
\end{equation}

Building on this relation, the characteristic Einstein radius corresponding to the characteristic VDF velocity dispersion can be expressed as
\begin{equation}  
\theta_{\rm E*} = \lambda(e)\left[4\pi \left(\frac{\sigma_{*\rm ref}}{c}\right)^2 \theta_{\rm ref}^{{\gamma-2}}  \frac{D_{ls}}{D_s} f(\gamma,\delta,\beta) \right]^{\frac{1}{\gamma-1}}. 
\label{eq:thetaE_scale}  
\end{equation}
Here $\sigma_{*\rm ref}$ is the characteristic velocity dispersion of the VDF evaluated at its reference radius. The factor $\lambda(e)$ accounts for the dynamical normalization of non-spherical galaxies \citep{1998ApJ...495..157K}, and $e$ denotes the projected ellipticity of the lens mass distribution, related to the projected minor-to-major axis ratio by $q = 1 - e$, consistent with the singular isothermal ellipsoid convention adopted by \citet{2012AJ....143..120O}. In this work, we adopt a fixed mean ellipticity $e=0.25$, consistent with the axis-ratio distributions of ETGs observed in the SDSS survey \citep{2010MNRAS.404.2087B}, and approximate the three-dimensional shapes of lens galaxies as an equal-weight combination of oblate and prolate isothermal spheroids:
\begin{equation}
\lambda(e) = 0.5\lambda_{\rm obl}(e) + 0.5\lambda_{\rm pro}(e),
\end{equation}
with
\begin{equation}
\begin{aligned}
\lambda_{\rm obl}(e) & \approx \exp\left(0.108\sqrt{e} + 0.180e^2 + 0.797e^5\right), \\
\lambda_{\rm pro}(e) & \approx 1 - 0.258e + 0.827e^6.
\end{aligned}
\end{equation}

Since the VDF of ETGs evolves with redshift, we adopt different sets of Schechter-function parameters $(\alpha,\beta^{\prime}, n_*, \sigma_*)$ for different redshift intervals, as summarized in Table~\ref{Table: VDF}. The VDF is written as
\begin{align}
\frac{dN}{d\sigma_{\rm ref}} &= n_*\left(\frac{\sigma_{\rm ref}}{\sigma_{*\rm ref}}\right)^\alpha 
\exp\left[-\left(\frac{\sigma_{\rm ref}}{\sigma_{*\rm ref}}\right)^{\beta^{\prime}}\right] \nonumber \\
&\quad \times \frac{\beta^{\prime}}{\Gamma(\alpha/\beta^{\prime})} \frac{1}{\sigma_{\rm ref}},
\label{eq:vdf_sigma}
\end{align}
which gives the comoving number density of galaxies per unit velocity dispersion. Here $\sigma_{\rm ref}$ denotes the stellar velocity dispersion defined at the same aperture convention as the adopted VDF, and $\sigma_{*\rm ref}$ is the corresponding characteristic velocity dispersion. The parameters $\alpha$ and $\beta^{\prime}$ control the low-velocity slope and the high-velocity cutoff of the VDF, respectively, while $n_*$ and $\sigma_{*\rm ref}$ represent the characteristic comoving number density and velocity dispersion.

Because the characteristic velocity dispersions adopted in different VDF measurements are either measured within or corrected to different aperture conventions, we assign a reference angular scale $\theta_{\rm ref}$ to each VDF dataset. For \citet{2007ApJ...658..884C}, stellar velocity dispersions were measured from SDSS spectra obtained with $3''$-diameter fibres and corrected to a standard aperture of $R_e/8$. To estimate the physical scale corresponding to this aperture correction, we use the early-type SDSS size--velocity-dispersion relation of \citet{2009MNRAS.394.1978H}, whose velocity dispersions are defined with the same $R_e/8$ aperture convention. For $\sigma_* = 161\,{\rm km\,s^{-1}}$, their $R_e$--$\sigma$ scaling relation gives $R_e \simeq 2.6\,{\rm kpc}$, and hence $R_e/8 \simeq 0.33\,{\rm kpc}$. At the representative lens redshift of the \citet{2007ApJ...658..884C} sample, $z_l = 0.07$, this gives $\theta_{\rm ref} \simeq 0.24''$. For \citet{2017MNRAS.468...47M}, we use their ``raw'' VDF, in which stellar velocity dispersions are measured directly within the BOSS $2''$-diameter spectroscopic fibre ($\theta_{\rm ref}=1.0''$), without any aperture correction. For \citet{2022ApJ...939...90T}, the LEGA-C stellar velocity dispersions are measured from spatially integrated one-dimensional spectra. Unlike a central small-aperture velocity dispersion, this quantity reflects the luminosity-weighted line broadening of the stellar light collected by the slit. The measured dispersion is therefore tied to the spatial distribution of the stellar light that dominates the integrated spectrum. As a characteristic scale of this main stellar light distribution, we adopt the projected half-light radius $R_e$ as the effective radius and set $\theta_{\rm ref}=\theta_{R_e}$. For quiescent galaxies at $z\sim0.8$, adopting a typical LEGA-C stellar mass of $\log(M_*/M_\odot)\simeq 10.75$ \citep{2018ApJS..239...27S}, the quiescent mass-size relation of \citet{2014ApJ...788...28V} gives $R_e \simeq 2.86\,{\rm kpc}$. At $z_l=0.8$, this corresponds to $\theta_{\rm ref}\simeq 0.38''$. We note that the systematic uncertainty associated with different aperture conventions is expected to be limited to the ${\sim}0.06\,{\rm dex}$ offset between $\sigma_{R_e}$ and $\sigma_{(R_e/8)}$ reported by \citet{2022ApJ...939...90T}, which remains smaller than the statistical uncertainties of the VDF Schechter parameters.

Returning to Eq.~(\ref{eq:dtao_dzl}), the comoving number density $n(\theta_{\rm E}, z_l)$ can be related to the VDF in Eq.~(\ref{eq:vdf_sigma}) via a change of variables,
\begin{equation}
n(\theta_{\rm E},z_l) d\theta_{\rm E}= \frac{dN}{d\sigma_{\rm ref}}(\sigma_{\rm ref},z_l)d\sigma_{\rm ref}.
\end{equation}
Using Eq.~(\ref{eq:thE_sigma}), the ratio $\sigma_{\rm ref}/\sigma_{*\rm ref}$ can be replaced by the corresponding Einstein-radius ratio,
\begin{equation}
R \equiv \frac{\sigma_{\rm ref}}{\sigma_{*\rm ref}}
= \left[ \frac{\lambda(e)\theta_{\rm E}}{\theta_{\rm E*}} \right]^{(\gamma-1)/2}.
\label{eq:R}
\end{equation}
Under this transformation, the differential optical-depth kernel evaluated at the observed Einstein radius becomes
\begin{equation}
\begin{aligned}
\frac{\boldsymbol{d \tau}}{\boldsymbol{d z_l}}\left( \boldsymbol{z_l|z_s,} \boldsymbol{\theta_{\rm E}}\right)
&= n_* \frac{({\gamma} - 1)\pi  \beta^{\prime}}{\Gamma(\alpha/\beta^{\prime})} \frac{D_l^2 (1+z_l)^2 c}{2 H(z_l)} \\
&\times {\theta}_{\rm E} R^{\alpha} \exp(-R^{ \beta^{\prime}}).
\label{eq:optical_depth}
\end{aligned}
\end{equation}

\begin{revised}

For a given strong lensing system $i$, the relative probability density of finding its lens galaxy at redshift $z_{l,i}$ is given by the differential optical depth of Eq.~(\ref{eq:optical_depth}) evaluated at $z_{l,i}$, normalized by its integral along the line of sight to the source. This normalization implicitly assumes that every lens system able to produce the observed configuration has the same probability of being recorded. In practice, a system that lies at $0<z_l<z_s$ and produces the observed Einstein radius $\theta_{\rm E}$ does not automatically enter our lens catalogue: it must first be identified as a lens and have both its lens and source redshifts measured. We therefore introduce a selection function $S$, defined as the probability that a system with given intrinsic properties is included in our compilation,
\begin{equation}
S = P\bigl(\,\mathrm{selected} \bigm| \mathrm{intrinsic\ properties}\,\bigr),
\label{eq:S_tex}
\end{equation}
so that the conditional likelihood can be written as
\begin{equation}
p_i(\mathbf{p}) = p(z_{l,i} \mid \theta_{E,i}, z_{s,i}, \mathbf{p}) = \frac{\left. \frac{d\tau}{dz_l}\bigl(\theta_{E,i}, z_l \bigm| z_{s,i}\bigr)\, S \right|_{z_l = z_{l,i}}}{\displaystyle \int_0^{z_{s,i}} \frac{d\tau}{dz'_l}\bigl(\theta_{E,i}, z'_l \bigm| z_{s,i}\bigr)\, S \, dz'_l}\, .
\label{eq:like_S}
\end{equation}
The integral in the denominator runs over the full foreground path to the source, $0<z_l<z_{s,i}$, following the standard lens-redshift test. Eq.~(\ref{eq:like_S}) therefore constrains the shape of the lens-redshift distribution at a given Einstein radius, not the absolute number of lenses.

\subsection{The lens selection function}

Whether a strong lens system ends up in our sample depends on how it was found. Systems with confirmed lensing features and with both $z_l$ and $z_s$ measured have entered existing samples through three main routes.

\textit{Image-based identification.} Here the system is first recognized from lensing configurations such as multiple images, lensed arcs, or Einstein rings. The image-based identification has been done in two main ways. The traditional approach is to search wide-field imaging for prominent arcs or rings, which requires the background source to be bright enough, the lensed images to be well resolved, and the light of the foreground galaxy not to swamp them. Early surveys such as SL2S and LSD identified candidates in this way and then obtained deep spectroscopy of each target individually on large telescopes, from which both $z_l$ and $z_s$ were measured. More recently, wide-area surveys \citep[e.g.][]{2025SCPMA..6829511S,2025arXiv250909798A,AGEL2022} have confirmed large numbers of candidates in imaging first and then cross-matching those candidates with existing SDSS or DESI spectra to obtain their redshifts and velocity dispersions. 

\textit{Spectroscopic identification.} Here the system is first flagged from the spectrum of the foreground galaxy. Within the magnitude limits of a systematic spectroscopic survey, a galaxy whose fibre spectrum shows emission lines from a higher-redshift object in addition to its own absorption features is selected as a candidate, provided that these lines stand out above the residual noise left once the foreground spectrum has been subtracted. Such candidates then trigger follow-up imaging, typically with HST, and are retained once the lensing configuration is confirmed. The classic SLACS, S4TM, BELLS, and BELLS GALLERY strong lensing samples were all assembled in this way.

\textit{Source-triggered discovery.} In a small number of cases the discovery starts from the source rather than from the lens: a bright background object picked up in a radio or submillimetre survey, such as MG2016+112 or SDP.81, is later recognized as lensed, and the foreground galaxy is identified only afterwards, entering the sample passively and with its own properties playing no role in the discovery.

For a single survey, the selection function can be constructed in detail. \citet{FerramiWyithe2024,FerramiWyithe2025}, for instance, modelled the detectability of SL2S lenses by folding the detection signal-to-noise ratio, the bright-image fraction, and the seeing-limited resolution into the probability of a system being found. They showed that these effects substantially reduce the number of detectable lenses and reshape the redshift and mass distributions of the observed sample. \citet{Sonnenfeld_2024} worked in a different direction, building an empirical model of the SLACS selection function to correct for the over-representation of galaxies with high velocity dispersions, and found that the mean total mass density slope of the parent early-type population differs from that of the lens sample by $\Delta\gamma\sim0.01$. Both approaches rely on a complete description of the imaging and of the search algorithm of one particular survey. The sample used in our work, however, is drawn from many sources with heterogeneous selection criteria, and for most of them no selection function has been given. Compiled samples of this kind will become more common as future surveys enlarge the available catalogues. We therefore need a strategy that handles the different selection effects system by system within a common framework.

\citet{CapeloNatarajan2007} tested the robustness of the lens-redshift test and found that lenses discovered through their background sources introduce no appreciable bias in the lens-redshift distribution, whereas for systems found through the foreground galaxy or its environment the selection function depends strongly on the lens redshift itself. This distinction between what is set by the source and what is set by the lens, together with the fact that the integral in Eq.~(\ref{eq:like_S}) runs over $z_l$ alone, leads us to write the selection function as
\begin{equation}
S \;=\; S_A(\theta_E,\, z_s)\; S_B(z_l\,;\,\theta_E,\, z_s).
\label{eq:SA_SB}
\end{equation}
Here $S_A$ collects the factors that stay fixed as the lens is moved along the line of sight, while $S_B$ carries the dependence on $z_l$. Since $\theta_{E,i}$ and $z_{s,i}$ are held fixed by the observations, Eq.~(\ref{eq:SA_SB}) simply states where the $z_l$ dependence of each factor resides. Factors that are set by both the source and the lens are split in the same way: the detectability of a lensed image, for example, depends on the brightness of the arc as well as on the contamination from the foreground light, and only the part of it that changes with $z_l$ is assigned to $S_B$.

$S_A$ therefore collects the factors that depend on the background source properties and govern whether the source is detected and recognized: the brightness and angular separation of the lensed images or arc; the luminosities of the background emission lines and whether they fall within the spectral coverage of the spectrograph; the alignment of the images with respect to the fibre aperture; the success rate of the source-redshift measurement; and the lens modeling requirement that the system show an identifiable lensing configuration in the available imaging. Magnification bias also belongs here. Although it alters the effective number density of sources, at the given conditional observables $(\theta_E,\, z_s)$ it is determined by the source and does not change with $z_l$.

$S_B$ acts through the foreground galaxy: whether it falls within the target magnitude window of its parent spectroscopic survey; the probability that its redshift and velocity dispersion are successfully measured under a fixed exposure strategy; and the suppression of the background arc or emission lines by the light of the foreground galaxy itself. Our isolation criterion and the $z_l$-dependent part of the lens-light subtraction operate through the same apparent brightness, and are grouped here as well.

Since $S_A$ does not depend on $z_l$, it can be taken outside the integral, and Eq.~(\ref{eq:like_S}) reduces to
\begin{equation}
p_i^{\rm corr}
= \frac{S_A\left[\dfrac{d\tau}{dz_l}\, S_B\right]_{z_{l,i}}}
{S_A\displaystyle\int_0^{z_{s,i}} \dfrac{d\tau}{dz'_l}\, S_B\, dz'_l}
= \frac{\left[\dfrac{d\tau}{dz_l}\, S_B\right]_{z_{l,i}}}
{\displaystyle\int_0^{z_{s,i}} \dfrac{d\tau}{dz'_l}\, S_B\, dz'_l} .
\label{eq:like_SB}
\end{equation}
%

The remaining redshift-dependent selection enters through $S_B$. For the heterogeneous compilation considered here, we model the foreground-dependent effects that can be characterized across the sample through the survey magnitude windows and foreground-light suppression. We assess the sensitivity of the inferred slopes to this model in Sect.~\ref{sec:Stability}. Based on the three discovery channels summarized above, we consider these two contributions to $S_B$ in turn.

\textit{The first is the observational brightness window.} Lenses identified spectroscopically enter the sample only when the foreground galaxy falls within the target-selection magnitude window $[m_{\rm bright},\, m_{\rm lim}]$ of its parent survey, and the same thresholds apply to lenses found in imaging whose redshifts or velocity dispersions come from a cross-match with a spectroscopic database. This window determines both whether a system is observed and whether it has a spectrum that provides a redshift and velocity dispersion to satisfy our criteria. For lenses that were followed up with dedicated deep spectroscopy, the exposure could be tailored to each target and integrated until both redshifts were measured, so no such window applies.

\textit{The second is the suppression of the background signal by the foreground light.} In spectroscopic identification the background emission line must stand above the residual noise left by the subtraction of the foreground spectrum, and that noise grows with the continuum flux of the lens galaxy. In image-based identification the lensed images must be separable from the foreground light which governs both whether the configuration is recognised as lensing and whether a candidate is convincing enough to be followed up spectroscopically. The two mechanisms are physically distinct, but both act in the same direction: the brighter the foreground galaxy, the harder the background signal is to detect.

What the two categories have in common is that both act through the apparent brightness of the foreground galaxy, $S_B$ is therefore governed by the apparent magnitude $m_{\rm lens}$ of the lens galaxy. As $m_{\rm lens}$ approaches or crosses either the faint-end target-selection limit $m_{\rm lim}$ or the bright-end saturation limit $m_{\rm bright}$, the probability that the system enters the sample drops. In turn, $m_{\rm lens}$ depends on the redshift of the lens, since it is determined by the rest-frame absolute magnitude $M_{\rm lens}(z_l)$, the distance modulus $\mu(z_l)$, the K-correction $K(z_l)$, and the evolutionary correction $E(z_l)$,
\begin{equation}
m_{\rm lens}(z_l) = M_{\rm lens}(z_l) + \mu(z_l) + K(z_l) + E(z_l).
\label{eq:mlens}
\end{equation}
For the absolute magnitude we follow the Faber-Jackson velocity-dispersion-magnitude relation of \citet{Bernardi_PaperII}, $M_{\rm lens}(z_l) = M_* + A\bigl(\log_{10}\sigma_{\rm ref}(z_l) - V_*\bigr)$, where $\sigma_{\rm ref}(z_l) = \sigma_{*\rm ref}\,R(z_l;\gamma,\delta,\beta)$ is the velocity dispersion the galaxy would need in order to produce the observed Einstein radius if it were placed at $z_l$, obtained from the ratio $R$ defined in Eq.~(\ref{eq:R}). The relation is calibrated on an SDSS early-type galaxy sample, in which the joint distribution of $(M,\log_{10}\sigma)$ is described by a bivariate Gaussian. Conditioning on the velocity dispersion gives $P\bigl(M \mid \log_{10}\sigma\bigr)=\mathcal{N}\!\bigl(M_* + A(\log_{10}\sigma - V_*),\, w^2\bigr)$, so that the absolute magnitude is normally distributed around $M_* + A(\log_{10}\sigma - V_*)$ with scatter $w$. Both the slope $A$ and the scatter $w$ are set by the covariance parameters of \citet{Bernardi_PaperII}, as detailed in Appendix~\ref{app:sigma_to_magnitude}. For the remaining terms in Eq.~(\ref{eq:mlens}), the $K$-correction removes the bandpass shift caused by redshift and converts the observed magnitude to a common rest-frame band, while $E$ is the passive luminosity-evolution correction, which accounts for the change in the intrinsic luminosity of the galaxy as its stellar population ages with redshift. The methods, templates, and numerical values used for these three terms are given in Appendix~\ref{app:absolute_to_apparent}.

From Eq.~(\ref{eq:mlens}), for a given velocity dispersion the absolute magnitude of the lens galaxy is not uniquely determined but follows a distribution. The apparent magnitude should therefore be written as $m_{\rm lens} + \epsilon$, where $\epsilon \sim \mathcal{N}(0, w^2)$ and $w$ is the width set by the intrinsic scatter of the Faber--Jackson relation. Whether a system falls within a given magnitude window is thus not a deterministic decision but a probability. For a window $m_{\rm bright} < m < m_{\rm lim}$ this probability is
\begin{equation}
P\big(m_{\rm bright} < m < m_{\rm lim}\big)
=
\Phi\!\left(\frac{m_{\rm lim}-m_{\rm lens}}{w}\right)
-
\Phi\!\left(\frac{m_{\rm bright}-m_{\rm lens}}{w}\right) ,
\label{eq:window}
\end{equation}
where $\Phi(\cdot)$ is the standard normal cumulative distribution function. $S_B$ can accordingly be written as the product of the brightness-window term and the foreground-light suppression term,
\begin{equation}
\begin{split}
S_B(z_l) \;=\;
&\underbrace{\left[\Phi\!\left(\frac{m_{\rm lim}-m_{\rm lens}}{w}\right)
-\Phi\!\left(\frac{m_{\rm bright}-m_{\rm lens}}{w}\right)\right]}_{\text{magnitude window}} \\[4pt]
\times\;
&\underbrace{\left[1-\varepsilon\,\Phi\!\left(\frac{m_{\rm sup}-m_{\rm lens}}{w}\right)\right]}_{\text{foreground suppression}} .
\end{split}
\label{eq:SB}
\end{equation}
The first term gives the probability that the lens galaxy is included in the spectroscopic survey. A galaxy that is too faint is unlikely to yield a usable spectrum, while one that is too bright may be rejected because of saturation or fibre cross-talk. Both limits are set by the published specifications of each parent survey. The second term accounts for the suppression of the background signal by the foreground light, which reduces the probability that the system is recognised as a lens or that a redshift is successfully extracted. Here $\varepsilon$ is the maximum loss in detection probability caused by this effect ($0\le\varepsilon\le1$), and $m_{\rm sup}$ is the brightness at which the suppression becomes appreciable. For lens magnitudes fainter than $m_{\rm sup}$ the factor tends to unity, and for brighter magnitudes it tends to $1-\varepsilon$. No external calibration is available for $m_{\rm sup}$, so we place it at the 10th percentile of the predicted magnitude distribution of the lens sample in each selection category. This location corresponds to the brightest systems in each category and quantifies the largest impact that foreground suppression can have. This factor is evaluated with the predicted $i$-band magnitude, as described in Appendix~\ref{app:SB_win}.


\end{revised}

\subsection{Statistical inference and MCMC sampling}

The normalization is performed over the full foreground path to the source, $0<z_l<z_s$, following the standard lens-redshift test. This form focuses the inference on the relative lens-redshift probability at the observed Einstein radius rather than on the absolute lensing rate. The total log-likelihood for the observed $N_{\rm lens}$ sample is then
\begin{equation}
\ln\mathcal{L}(\mathbf{p}) =
\sum_{i=1}^{N_{\rm lens}} \ln p_i^{\rm corr}(\mathbf{p}),
\label{eq:total_likelihood}
\end{equation}
where the model parameter vector is
\begin{equation}
\mathbf{p}=(\gamma,\delta,\beta).
\end{equation}
We employ a Bayesian Markov Chain Monte Carlo (MCMC) framework to explore the posterior distributions of the model parameters. For each redshift interval and VDF choice, we use 32 walkers and 50000 steps, discarding the first 20\% of each chain as burn-in. Convergence is assessed via the Gelman--Rubin statistic and the integrated autocorrelation time $\tau$; all chains satisfy $\hat{R}<1.01$ and $N_\mathrm{kept}/\tau > 50$, confirming adequate mixing and convergence. The reported constraints correspond to the median and the 16th-84th percentile credible intervals of the marginalized posterior distributions.

The analysis is performed independently in the three redshift intervals defined by the available VDF measurements. For the fiducial case, we use the central VDF parameters in each interval. To quantify the impact of VDF uncertainties and sample construction, we perform a series of targeted robustness tests to assess the stability of the inferred density slopes. The resulting posterior samples and marginalized likelihood contours serve as the basis for the statistical analysis presented in Section~\ref{sec:results}.

\section{Observational data} 
\label{sec:Data}

\subsection{Compilation of the lens sample}

\begin{figure}
  {{\includegraphics[width=3.2 in]{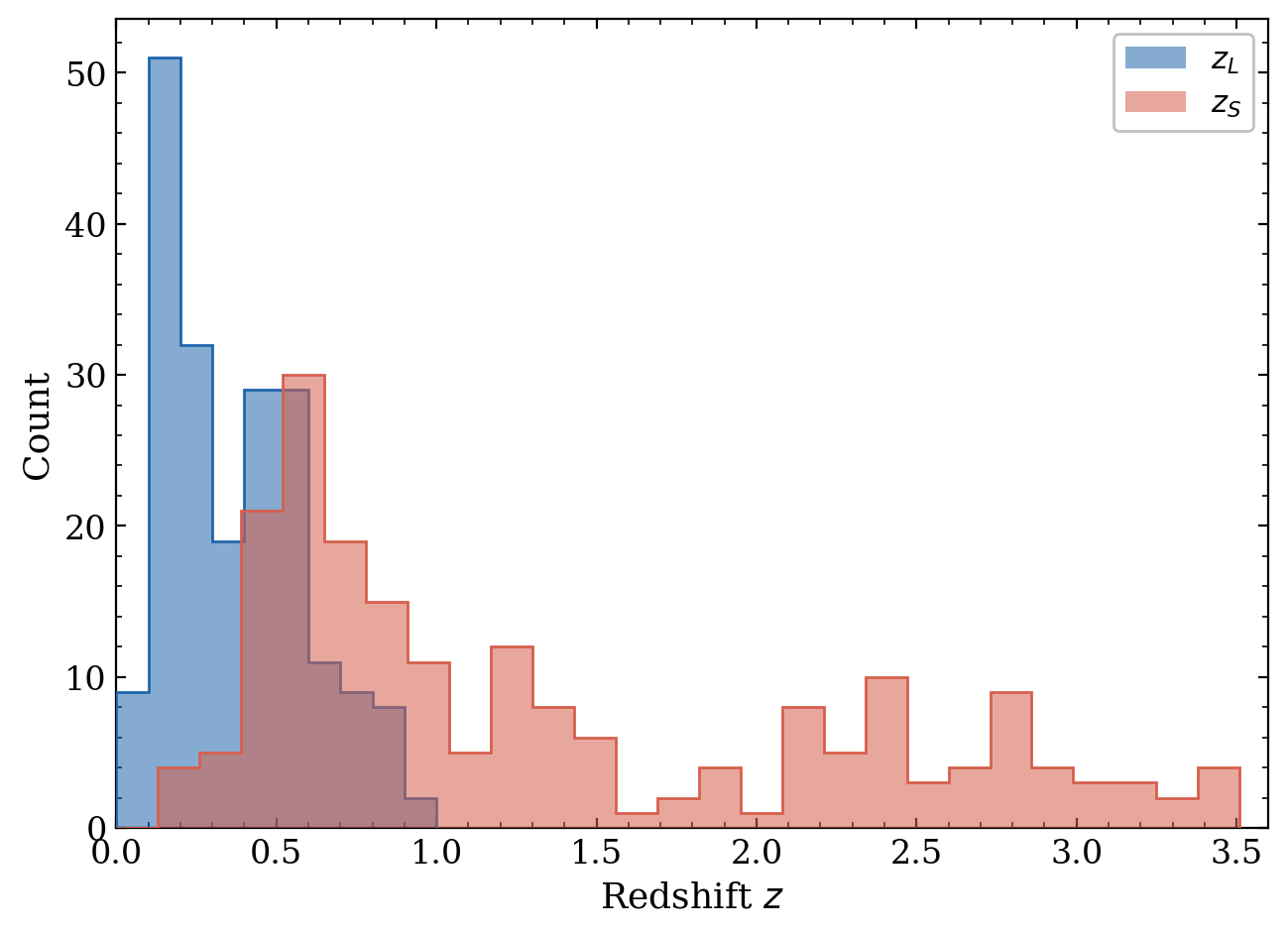}}}
  \caption{Redshift distribution of the strong lensing sample.}
  \label{fig:z_distributions}
\end{figure}

\begin{figure}
  {{\includegraphics[width=3.2 in]{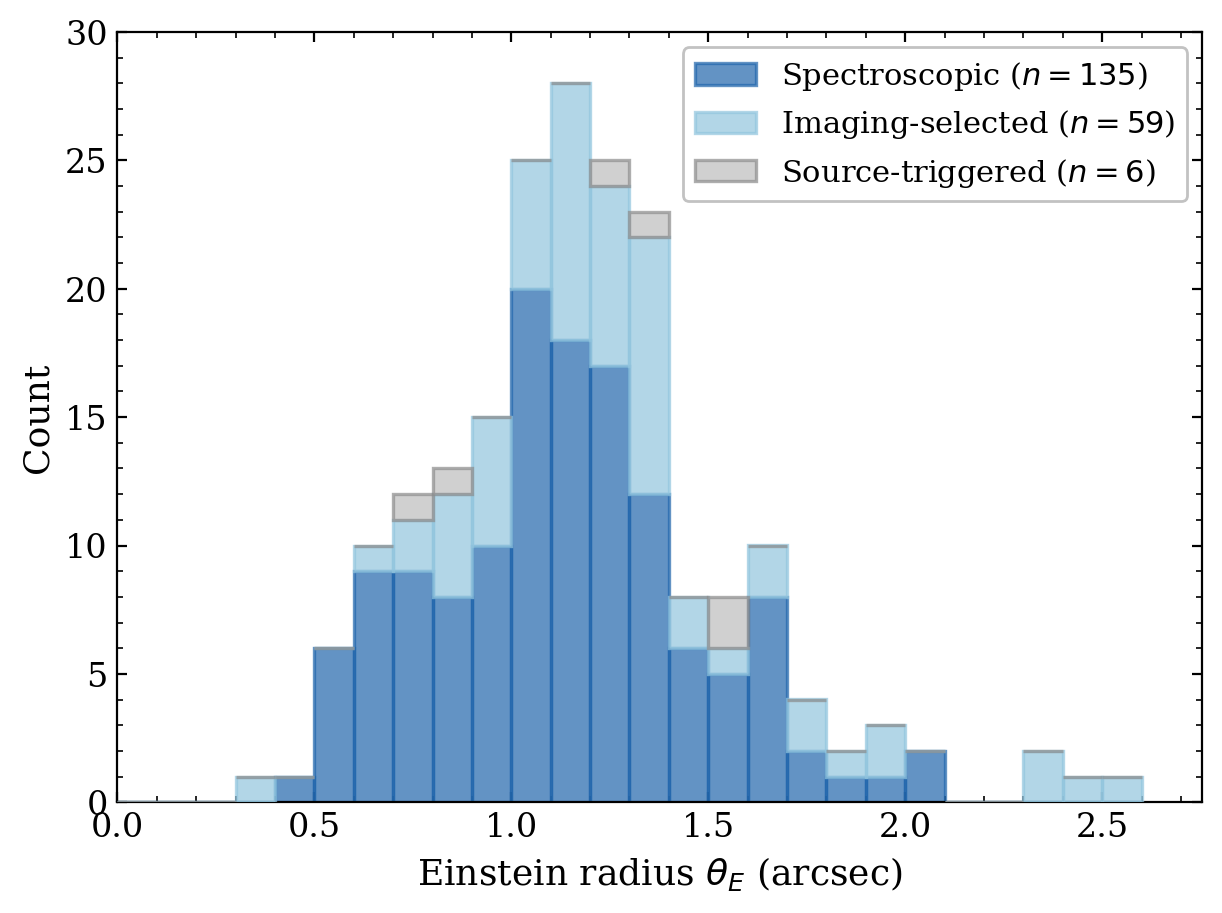}}}
  \caption{Einstein radius distribution of the strong lensing sample. The histograms show lenses discovered through spectroscopy ($n=135$), through lens imaging ($n=59$), and through source-triggered detections ($n=6$).}
  \label{fig:thE_distributions}
\end{figure}

Our analysis is based on a newly compiled sample of 200 galaxy-scale strong gravitational lens systems. Motivated by known systematic differences between lens galaxies in group/cluster environments and isolated lenses \citep[][]{2010ApJ...724..511A}, we focus on galaxy-scale early-type lenses that appear isolated based on currently available imaging and literature information. Lens galaxies reported as disky or late-type in the discovery papers or high-resolution imaging are excluded. To be included in the sample, each system must exhibit identifiable lensing features in available imaging and have the observational quantities required by our statistical framework: lens redshift $z_l$, source redshift $z_s$, Einstein radius $\theta_E$, spectroscopic aperture radius $\theta_\mathrm{ap}$, and aperture velocity dispersion $\sigma_\mathrm{ap}$. The velocity-dispersion information is used to define a kinematically characterized lens sample and to enable data-quality checks, while the baseline statistical likelihood is constructed from the conditional optical-depth formalism described in Sect.~\ref{sec:Meth}. We emphasize that the statistical likelihood does not use the individual $\sigma_{\rm ap}$ measurements as lens-by-lens dynamical constraints; instead, the kinematic information enters the inference statistically through the VDF of the deflector population. For each system we also record the channel through which it was discovered and the spectroscopic survey from which its redshift and velocity dispersion come, since these set the selection-function parameters introduced in Sect.~\ref{sec:Meth} (see Sect.~\ref{sec: data_sel}).

\begin{figure*}[t!]
\centering
\includegraphics[width=0.97\textwidth]{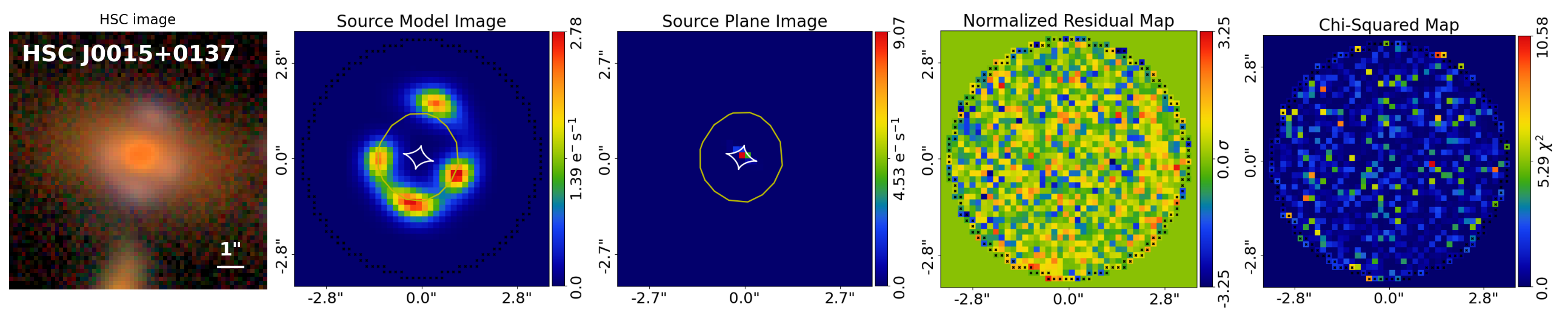}\\[2pt]
\includegraphics[width=0.97\textwidth]{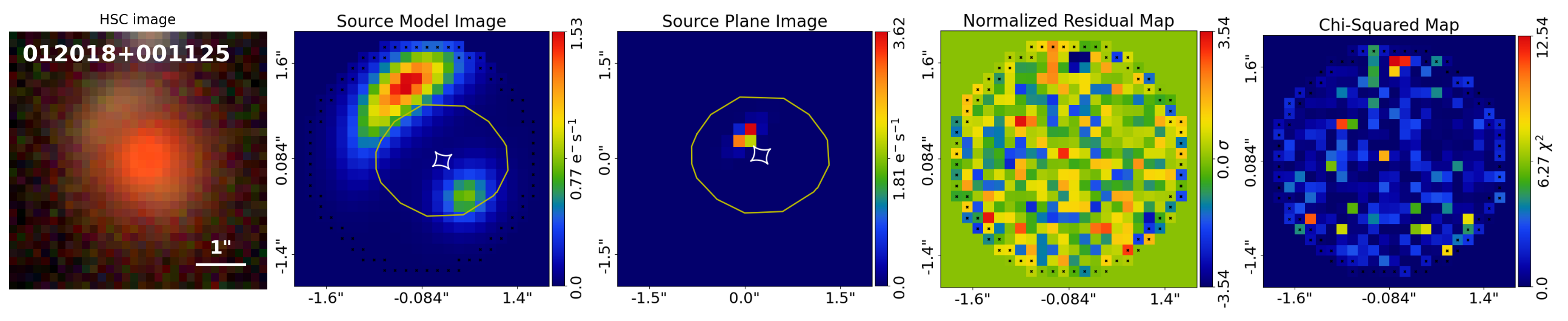}\\[2pt]
\includegraphics[width=0.97\textwidth]{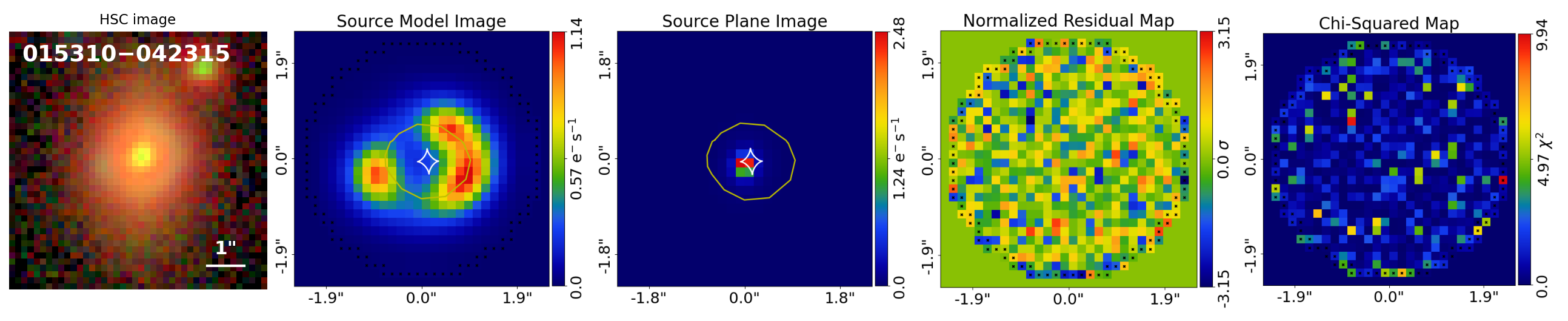}\\[2pt]
\includegraphics[width=0.97\textwidth]{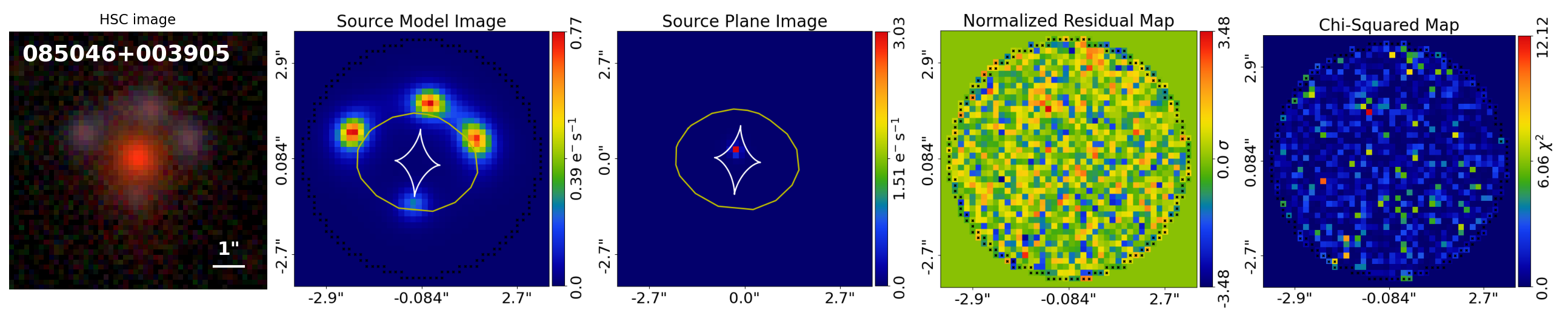}\\[2pt]
\includegraphics[width=0.97\textwidth]{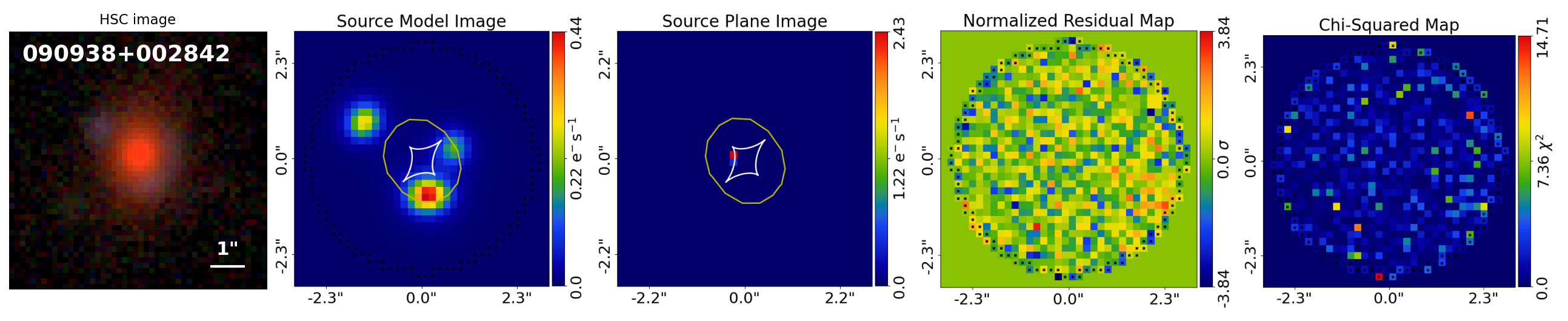}\\[2pt]
\includegraphics[width=0.97\textwidth]{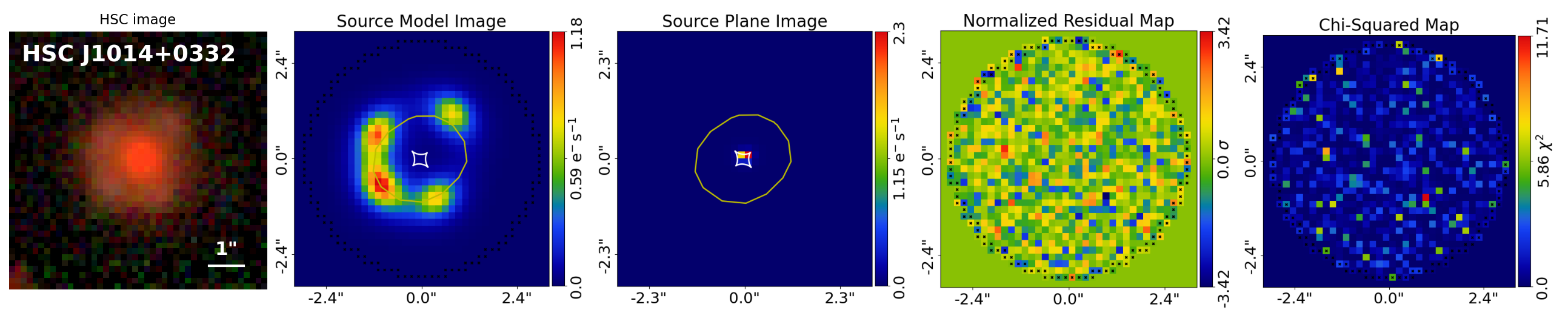}
\caption{Lens modeling results for the 13 strong gravitational lens
systems modeled in this work. Each row shows, from left to right: the
observed HSC image, the reconstructed lensed source model, the
source-plane reconstruction, the normalized residual map (in units of
$\sigma$), and the $\chi^2$ map. The system name is labeled in white
in the leftmost panel of each row.}
\label{fig:lens_modeling}
\end{figure*}

\begin{figure*}[t!]
\ContinuedFloat
\centering
\includegraphics[width=0.97\textwidth]{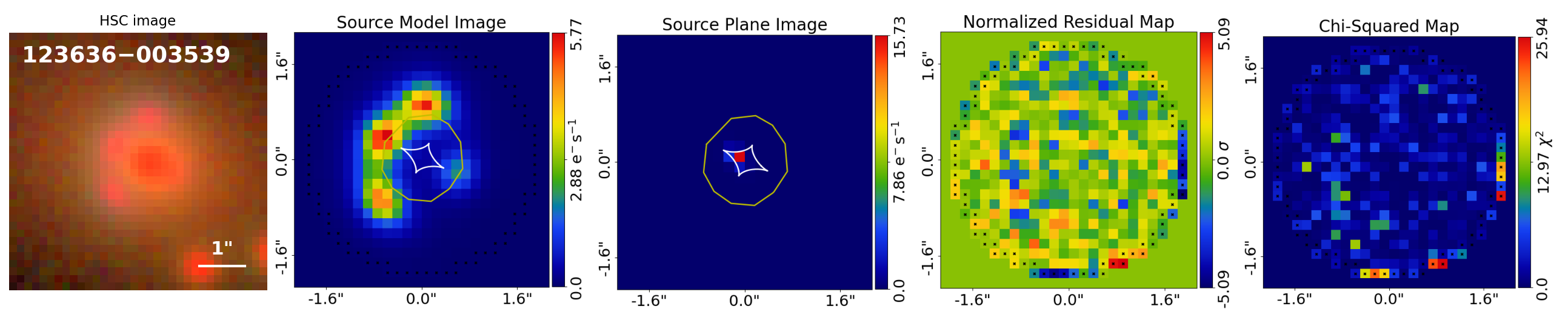}\\[2pt]
\includegraphics[width=0.97\textwidth]{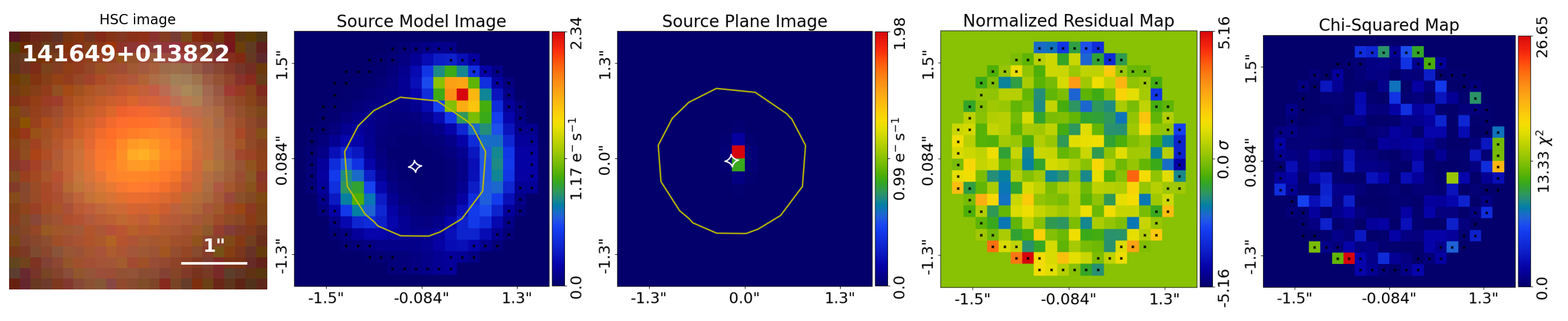}\\[2pt]
\includegraphics[width=0.97\textwidth]{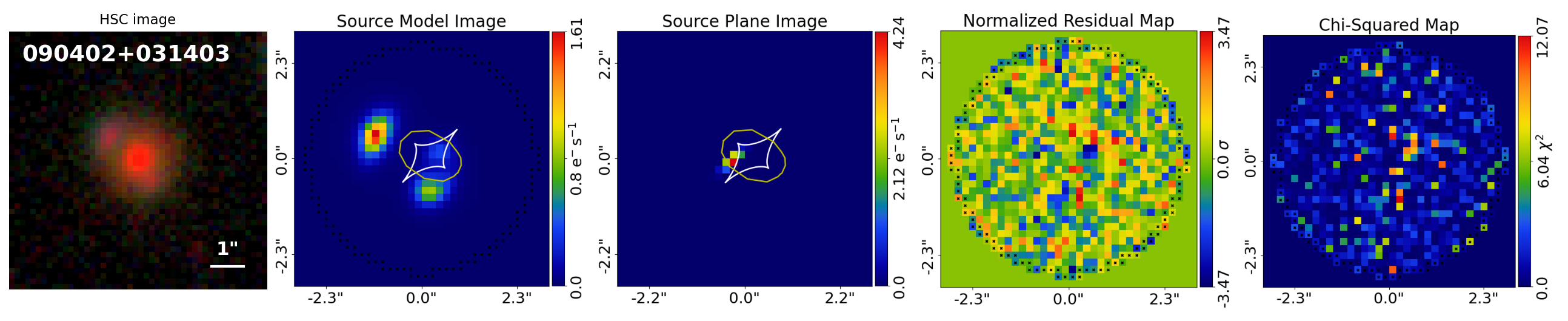}\\[2pt]
\includegraphics[width=0.97\textwidth]{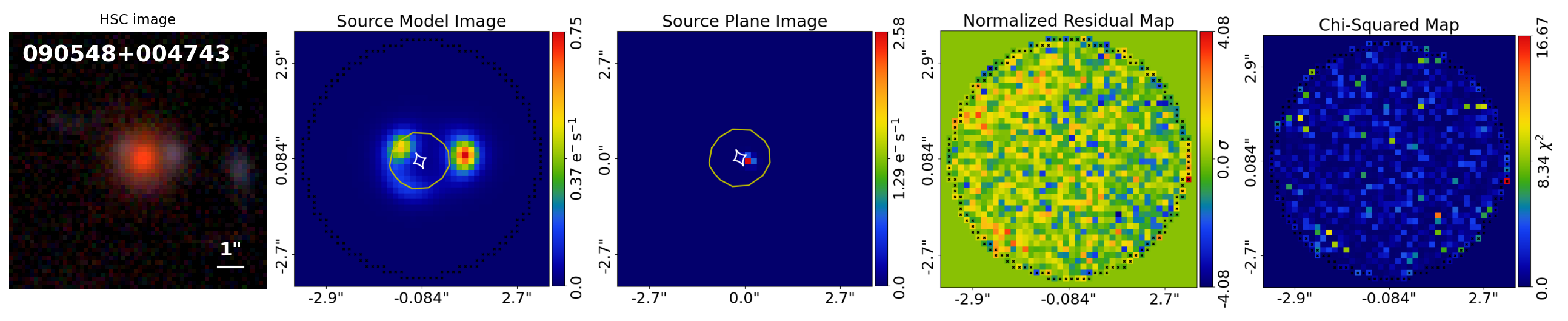}\\[2pt]
\includegraphics[width=0.97\textwidth]{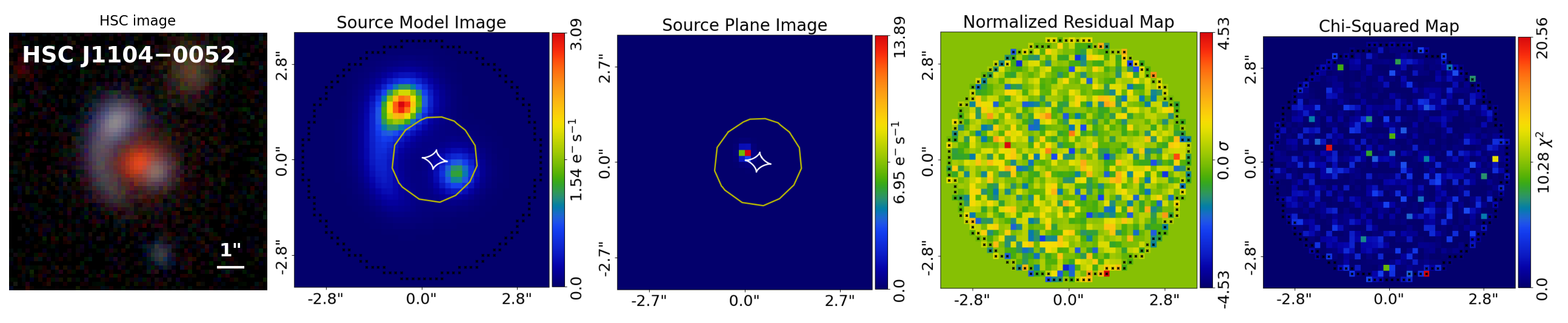}\\[2pt]
\includegraphics[width=0.97\textwidth]{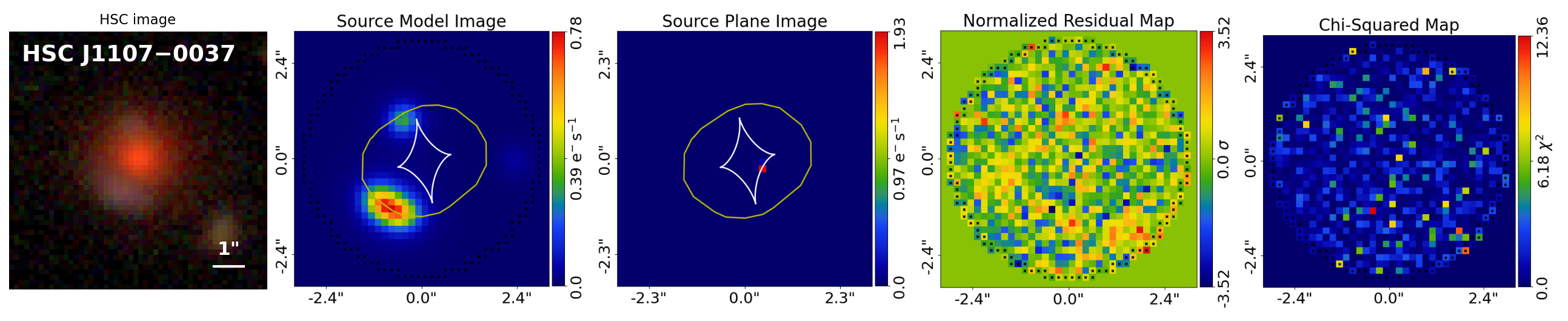}
\caption{continued.}
\end{figure*}

\begin{figure*}[t!]
\ContinuedFloat
\centering
\includegraphics[width=0.97\textwidth]{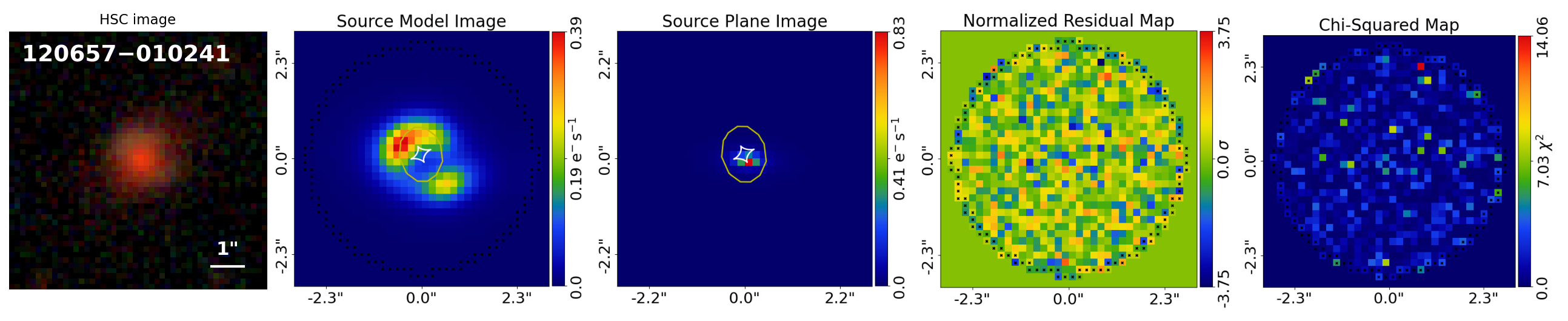}
\caption{continued.}
\end{figure*}

\newlength{\mytabcolsep}
\setlength{\mytabcolsep}{6pt}   

\begin{table*}
\centering
\caption{Strong lens systems newly added on top of \citet{2019MNRAS.488.3745C}, drawn from \citet{2025arXiv250909798A}, \citet{Guo2026}, recent surveys and our own lens modelling. Columns list the system name, lens redshift $z_l$, source redshift $z_s$, Einstein radius $\theta_E$, angular aperture radius $\theta_\mathrm{ap}$ for velocity dispersion, aperture-averaged velocity dispersion $\sigma_\mathrm{ap}$ with its uncertainty, and the data reference. The listed velocity dispersions characterize the spectroscopic properties of the lens galaxies and are not used as individual dynamical likelihood terms in the baseline statistical analysis.}
\label{tab:new_sample}
\setlength{\tabcolsep}{\mytabcolsep}
\begin{tabular}{lcccccl}
\hline\hline
Name & $z_l$ & $z_s$ & $\theta_E$ & $\theta_\mathrm{ap}$ & $\sigma_\mathrm{ap}$ & Ref. \\
     &        &        & (arcsec)   & (arcsec)             & (km\,s$^{-1}$)       &      \\
\hline
\multicolumn{7}{l}{\textit{From Alves de Oliveira et al. (2025)}} \\
\hline
J085559.0$-$040917.0 & 0.4190 & 2.950 & 1.36 & 0.70 & $281\pm22$   & \citet{2013ApJ...777...97S} \\
J020241.4$-$064611.3 & 0.5020 & 2.748 & 1.16 & 1.00 & $160\pm26$   & \citet{2024MNRAS.535.1625J} \\
J023307.1$-$043838.2 & 0.6709 & 1.869 & 1.77 & 1.00 & $386\pm111$  & \citet{refId0} \\
J023817.8$-$054555.6 & 0.5993 & 1.763 & 0.93 & 1.00 & $382\pm86$   & \citet{2024MNRAS.527.6253C} \\
J090311.6$+$003906.5 & 0.2997 & 3.042 & 1.56 & 1.00 & $243\pm16$   & \citet{2015ApJ...808L...4A} \\
J103121.0$+$302658.0 & 0.1671 & 0.747 & 0.88 & 1.50 & $196\pm13$   & \citet{2017ApJ...851...48S} \\
J111040.4$+$364924.4 & 0.7330 & 2.500 & 1.99 & 1.00 & $354\pm116$  & SuGOHI$^{a}$ \\
J120623.9$+$001507.1 & 0.5632 & 3.120 & 1.17 & 1.00 & $271\pm47$   & \citet{refId0} \\
J121052.5$-$011905.2 & 0.6998 & 2.295 & 1.16 & 1.00 & $334\pm66$   & \citet{refId0} \\
J140929.7$-$011410.8 & 0.5842 & 2.302 & 1.24 & 1.00 & $187\pm33$   & \citet{refId0} \\
J141300.1$-$012608.2 & 0.7490 & 2.666 & 1.14 & 1.00 & $243\pm67$   & \citet{refId0} \\
J141815.7$+$015832.3 & 0.5559 & 2.139 & 1.40 & 1.00 & $219\pm30$   & \citet{refId0} \\
J143213.3$+$631703.8 & 0.1231 & 0.664 & 1.26 & 1.50 & $194\pm8$    & \citet{2008ApJ...682..964B} \\
J143607.5$+$494313.2 & 0.1254 & 1.231 & 2.03 & 1.50 & $282\pm10$   & \citet{2020MNRAS.493L..33S} \\
\hline
\multicolumn{7}{l}{\textit{Galaxy--quasar systems}} \\
\hline
SDSSJ0114$+$0722   & 0.4080 & 1.828 & 0.830 & 0.75 & $222\pm20$ & \citet{2016MNRAS.456.1595M} \\
J0203$+$1612       & 0.4880 & 2.180 & 1.360 & 1.00 & $207\pm50$ & \citet{Lemon2019} \\
HSCJ0918$-$0220    & 0.4600 & 0.803 & 1.280 & 0.75 & $360\pm30$ & \citet{Jaelani2021} \\
J0949$+$4208       & 0.5100 & 1.270 & 1.230 & 0.75 & $290\pm50$ & \citet{2018MNRAS.479.5060L} \\
COSMOS5921$+$0638  & 0.5510 & 3.140 & 0.700 & 0.75 & $186\pm30$ & \citet{Faure2008} \\
HSCJ1215$-$0058    & 0.4591 & 2.882 & 2.342 & 0.75 & $220\pm60$ & \citet{He2025} \\
HSCJ1220$+$0112    & 0.4876 & 1.708 & 0.958 & 0.75 & $222\pm40$ & \citet{He2025} \\
SDSSJ1524$+$4409   & 0.3200 & 1.210 & 0.790 & 0.75 & $222\pm15$ & \citet{2008AJ....135..520O} \\
SDSSJ1640$+$1932   & 0.1950 & 0.778 & 2.490 & 1.00 & $373\pm8$  & \citet{Wang2017} \\
\hline
\multicolumn{7}{l}{\textit{From recent surveys}} \\
\hline
2MASXJ04035024$-$0239275 & 0.0660 & 0.192 & 1.470 & 1.50 & $314\pm5$  & \citet{Collier2018} \\
AGEL053724$-$464702      & 0.3520 & 2.344 & 1.902 & 0.66 & $288\pm19$ & \citet{2024ApJ...970...86S} \\
AGEL233610$-$020735      & 0.4942 & 2.662 & 1.392 & 1.00 & $272\pm35$ & \citet{2024ApJ...970...86S} \\
HSCJ0904$-$0102          & 0.9570 & 3.403 & 1.230 & 0.50 & $259\pm56$ & \citet{Jaelani2020} \\
\hline
\multicolumn{7}{l}{\textit{This work (modeled from Shu et al. 2025)}} \\
\hline
HSC J0015$+$0137  & 0.3788 & 1.165 & 1.344 & 0.75 & $272\pm21$ & \citet{2025SCPMA..6829511S} \\
012018$+$001125   & 0.5985 & 2.999 & 1.101 & 0.75 & $233\pm24$ & \citet{2025SCPMA..6829511S} \\
015310$-$042315   & 0.2873 & 0.648 & 0.895 & 0.75 & $338\pm37$ & \citet{2025SCPMA..6829511S} \\
085046$+$003905   & 0.8844 & 2.838 & 1.812 & 0.75 & $394\pm77$ & \citet{2025SCPMA..6829511S} \\
090938$+$002842   & 0.7795 & 2.546 & 1.094 & 0.75 & $387\pm65$ & \citet{2025SCPMA..6829511S} \\
HSC J1014$+$0332  & 0.6777 & 2.165 & 1.265 & 0.75 & $261\pm31$ & \citet{2025SCPMA..6829511S} \\
123636$-$003539   & 0.5102 & 0.763 & 0.819 & 0.75 & $370\pm32$ & \citet{2025SCPMA..6829511S} \\
141649$+$013822   & 0.4337 & 1.313 & 1.121 & 0.75 & $256\pm19$ & \citet{2025SCPMA..6829511S} \\
090402$+$031403   & 0.8468 & 3.353 & 0.854 & 0.75 & $177\pm26$ & \citet{2025SCPMA..6829511S} \\
090548$+$004743   & 0.8535 & 2.912 & 0.936 & 0.75 & $249\pm34$ & \citet{2025SCPMA..6829511S} \\
HSC J1104$-$0052  & 0.8730 & 2.456 & 1.337 & 0.75 & $455\pm91$ & \citet{2025SCPMA..6829511S} \\
HSC J1107$-$0037  & 0.8144 & 1.411 & 1.659 & 0.75 & $268\pm47$ & \citet{2025SCPMA..6829511S} \\
120657$-$010241   & 0.8300 & 2.773 & 0.756 & 0.75 & $193\pm45$ & \citet{2025SCPMA..6829511S} \\
\hline
\end{tabular}
\begin{minipage}{\linewidth}
  \smallskip
  \small
  $^{a}$ System information retrieved from the SuGOHI online database:
  {\url{https://www-utap.phys.s.u-tokyo.ac.jp/~oguri/sugohi/}}\\
\end{minipage}
\end{table*}

The foundation of our sample is the compilation presented in \citet{2019MNRAS.488.3745C}, which contains 161 galaxy-scale strong lens systems. One of them, SL2SJ095921$+$020638, is listed below among the galaxy-quasar systems under its \citet{Guo2026} designation, leaving 160 systems counted from this source. We further supplement the sample by drawing from the compilation of \citet{2025arXiv250909798A}. Rather than adopting their catalogue wholesale, we trace each system individually back to its original discovery reference and verify that it satisfies our selection criteria. Systems that lack a reliable primary source, show evidence of a group or cluster environment, or are identified as non-isolated are excluded. The velocity-dispersion values for these systems, which are useful for characterizing the lens galaxies and for consistency checks with dynamical analyses, are adopted from the SDSS cross-match reported in \citet{2025arXiv250909798A}. After screening, 14 systems are retained, with the accepted systems and their original references listed in Table~\ref{tab:new_sample}. As discussed in Sect.~\ref{sec:Meth}, the source properties do not affect the inference. We therefore also draw galaxy-quasar systems from \citet{Guo2026}. Applying the same criteria as above, we include these galaxy-quasar systems: SDSSJ0114$+$0722, J0203$+$1612, HSCJ0918$-$0220, J0949$+$4208, HSCJ1215$-$0058, HSCJ1220$+$0112, SDSSJ1524$+$4409, and SDSSJ1640$+$1932. Note that COSMOS5921$+$0638 in \citet{Guo2026} is the same system as SL2SJ095921$+$020638 in \citet{2019MNRAS.488.3745C}; it is counted once, in the galaxy-quasar block of Table~\ref{tab:new_sample}, where we adopt the updated measurements of \citet{Guo2026}. In addition, we include four systems from recent new surveys that satisfy the same criteria, namely 2MASXJ04035024$-$0239275, AGEL053724$-$464702, AGEL233610$-$020735, and HSCJ0904$-$0102, which extend the sample towards higher lens redshifts and larger Einstein radii.

We additionally perform independent lens modeling for systems drawn from \citet{2025SCPMA..6829511S}, who cross-matched strong lens candidates identified in the Hyper Suprime-Cam Subaru Strategic Program \citep[][]{2018PASJ...70S...4A} with DESI DR1 spectroscopy, confirming spectroscopic redshifts for both lens and source. DESI is conducting one of the largest spectroscopic surveys to date \citep{2016arXiv161100036D}, targeting tens of millions of galaxies and quasars with the primary goal of constraining the cosmic expansion history. Its wide-field spectroscopic capability has also proven effective for the confirmation of strong gravitational lens systems. We apply our selection criteria to their sample and model the HSC images using \textsc{PyAutoLens} \citep{2021JOSS....6.2825N,2024MNRAS.532.2441H}, primarily in the $i$ band, which generally provides the best balance between image quality and lens-galaxy light tracing for strong lensing modeling. For a small number of systems where the $i$-band imaging was of lower quality or did not yield a converged model, alternative bands were adopted: \object{015310-042315} in $g$, \object{141649+013822} in $r$, \object{HSC J1107-0037} in $g$, and \object{120657-010241} in $r$. The lens light is modeled with a multi-Gaussian expansion \citep[MGE;][]{1994A&A...285..723E} comprising two groups of co-centred Gaussians with logarithmically spaced widths and linearly solved intensities. The lens mass is modeled as a singular isothermal ellipsoid (SIE) with external shear, as is common in the literature, and the background source is modeled using the \textsc{SersicCore} light profile in \textsc{PyAutoLens}, which implements a softened Sérsic profile with a finite core radius to improve numerical stability. Since $\theta_E$ is expected to be a model-independent lensing quantity \citep{2022MNRAS.517.3275E}, the use of our lens model provides a robust estimate of the Einstein radius for the statistical analysis. Systems are retained only when the lens light subtraction is clean, the reconstructed source is physically plausible, and the lens model converges to a well-defined solution, as assessed from the residual maps and $\chi^2$ maps presented in Figure~\ref{fig:lens_modeling}. The velocity dispersions for these systems are extracted from the DESI DR1 \texttt{FastSpecFit} Spectral Synthesis and Emission-Line Catalogue\footnote{\url{https://data.desi.lbl.gov/public/dr1/vac/dr1/fastspecfit/iron/v2.1/catalogs/fastspec-iron.fits} and its \href{https://data.desi.lbl.gov/doc/releases/dr1/vac/fastspecfit/}{website}}, which provides derived spectrophotometric quantities including stellar velocity dispersions obtained from spectral fitting. For the DESI systems, the angular aperture radius is set by the DESI fibre radius, $\theta_\mathrm{ap}=0.75''$. These velocity dispersions and aperture radii provide a homogeneous spectroscopic characterization of the newly modeled subsample and are used consistently in compiling the observational quantities listed in Table~\ref{tab:new_sample}.

In total, 13 systems pass all criteria, with modeling reconstructions, residual maps, and $\chi^2$ maps presented in Fig.~\ref{fig:lens_modeling}. The final sample comprises systems from \citet{2019MNRAS.488.3745C}, \citet{2025arXiv250909798A}, the modelled systems from \citet{2025SCPMA..6829511S}, galaxy-quasar systems from \citet{Guo2026}, and additional systems from recent surveys, yielding a total of 200 systems. By combining legacy compilations with modern wide-field discoveries, our sample covers a lens redshift range of \(0.0625\leq z_l\leq1.004\) and an Einstein radius range of $0.36^{\prime\prime} \leq \theta_E \leq 2.55^{\prime\prime}$. This broad coverage supports a systematic investigation of the total mass-density slope in ETG lenses over a wide redshift baseline. As discussed above, lens systems reached our sample through three discovery channels. Spectroscopically selected lenses, from surveys such as SLACS, BELLS, and BELLS GALLERY ($n=135$), are identified through the detection of emission lines from background sources in the spectra of foreground galaxies. Image-selected lenses, from surveys such as SL2S, AGEL, and SuGOHI ($n=59$), are identified photometrically through arc or ring signatures in wide-field imaging. Source-triggered systems ($n=6$) were first noticed as bright background objects and only later recognized as lensed. The distributions of $z_l$, $z_s$, and $\theta_E$ are shown in Figs.~\ref{fig:z_distributions} and \ref{fig:thE_distributions}.

\subsection{Sample Selection Classes}
\label{sec: data_sel}

\begin{table*}
\centering
\caption{Summary of selection-function categories. The first column lists the discovery path and follow-up confirmation of each system; the second column gives the selection-function category identifier used in this work; third column $n$ is the number of systems; the fourth column lists the magnitude window used for target selection in each survey for specific band. Here $r_{\rm Petro}$ is the SDSS Petrosian magnitude used for the MAIN and LRG samples; $r_{\rm cmodel}$ and $i_{\rm cmodel}$ are the BOSS cmodel magnitudes used for the LOWZ and CMASS samples; and $z_{\rm fibre}$ is the DESI fibre magnitude in the $z$ band, providing the faint-end limit for LRG targeting. These magnitude definitions belong to different photometric systems and are not homogenized. When computing $S_B$, each magnitude is converted to the corresponding total magnitude system to match the lens apparent magnitude predicted by the Faber--Jackson relation; details of the conversion and adopted values are given in Appendix~\ref{app:SB_win}. Six systems in the sample were first noticed through the background source; the four whose redshifts and velocity dispersions come from an existing spectroscopic library are assigned to the category of that library, so the last row contains only the two for which they come from dedicated observations.}
\label{tab:selection_categories}
\setlength{\tabcolsep}{6pt}
\footnotesize
\begin{tabular}{llcl}
\hline\hline
Sample source and follow-up & Selection function category & $n$ & Magnitude selection \\
\hline \\
SDSS spectroscopic parent sample & SPEC\_SDSS\ (MAIN, LRG, SPECIAL) & 98 & $r_{\rm Petro} \in [14.5,\ 19.5]$ \\[5pt] 
BOSS spectroscopic parent sample & SPEC\_BOSS\ (CMASS, LOWZ) & 45 & $r_{\rm cmodel} \in [16.0,\ 19.6], i_{\rm cmodel} \in [17.5,\ 19.9]$ \\[5pt]
DESI spectroscopic parent sample & SPEC\_DESI & 13 & $z_{\rm fibre} < 21.60$ \\ [5pt]
Imaging confirmation + spectroscopic follow-up & IMAGING\_SPEC\_FOLLOWUP & 42 & None (foreground suppression only) \\[5pt]
Source-triggered + spectroscopic follow-up & SOURCE\_SPEC\_FOLLOWUP & 2 & None ($S_B \equiv 1$) \\[5pt]
\hline\hline
\end{tabular}
\end{table*}

Having assembled the sample from the various discovery channels described above, we now specify how the selection-function parameters introduced in Sect.~\ref{sec:Meth} are assigned to individual systems when computing $S_B$. The assignment follows the spectroscopic survey from which a system obtained its redshift and velocity dispersion, rather than only the way in which it was first noticed, since it is the parent spectroscopic sample that imposes the magnitude window. For systems targeted by a systematic spectroscopic survey, and for systems found in imaging whose spectroscopic quantities come from a cross-match with such a survey, we adopt the target-selection thresholds of that survey as $m_{\rm lim}$ and $m_{\rm bright}$. For systems followed up with dedicated spectroscopy of individual targets no such window applies, and only the foreground-suppression factor acts. For source-triggered systems the discovery does not depend on the foreground galaxy, so $S_B \equiv 1$. The foreground-suppression amplitude is applied uniformly to both the spectroscopically targeted and imaging-discovered categories, with a fiducial value of $\varepsilon = 0.2$. Since there is no specific external calibration for $\varepsilon$, we later vary it as a stress test of the assumed foreground-suppression strength. For the same reason, $m_{\rm sup}$ is set separately for each category to assess the maximum impact of foreground suppression in that category, as described in Sect.~\ref{sec:Meth}. The categories and the adopted magnitude windows are summarized in Table~\ref{tab:selection_categories}.

The rows labelled by the SDSS parent sample include both lenses spectroscopically confirmed by SDSS itself and systems discovered in imaging whose velocity dispersions come from a cross-match with SDSS. This category comprises three targeting programmes, labelled \texttt{MAIN}, \texttt{LRG}, and \texttt{SPECIAL}, the last referring to the SDSS ancillary programmes. Although the MAIN ($r_{\rm Petro}<17.77$) and LRG ($r_{\rm Petro}<19.5$) thresholds differ, a galaxy that fails the MAIN cut can still enter the LRG sample if it satisfies the LRG threshold, and a number of systems in our sample satisfy both criteria. Some targets from the ancillary programmes are fainter than $19.5$, the faintest of them exceeding this limit only marginally, with two at $19.8$. We adopt for them the unified proxy threshold of the SDSS spectroscopic parent sample, $r_{\rm Petro}<19.5$, and label them as \texttt{SPECIAL} for traceability. The bright-end limits quoted in Table~\ref{tab:selection_categories} are the saturation limits of the corresponding surveys. The BOSS parent-sample rows similarly include lenses spectroscopically confirmed by BOSS and imaging-discovered systems with velocity dispersions from a BOSS cross-match. The BOSS spectroscopic parent sample includes galaxies targeted through either \texttt{LOWZ} or \texttt{CMASS}. We approximate their combined magnitude selection at each trial redshift by the larger of the two window probabilities. Both the \texttt{SPEC\_SDSS} and \texttt{SPEC\_BOSS} categories therefore also contain imaging-discovered systems, such as some SuGOHI lenses, whose spectroscopic quantities are obtained by cross-matching. The \texttt{SPEC\_DESI} category is dominated by systems discovered in HSC imaging, with lens redshifts and velocity dispersions taken from the DESI spectroscopic library. Since the DESI LRG target selection is expressed as a faint-end limit within a fibre aperture, no bright-end truncation is applied in this category. \texttt{IMAGING\_SPEC\_FOLLOWUP} consists of systems discovered in imaging and subsequently followed up with dedicated spectroscopy of individual targets, from which their redshifts and velocity dispersions are obtained. With no parent spectroscopic survey imposing a magnitude window, this category is subject to the foreground-suppression factor $\varepsilon$ alone. \texttt{SOURCE\_SPEC\_FOLLOWUP} refers to systems that were first noticed through the background source and whose spectroscopic quantities come from dedicated observations, namely MG2016$+$112 and SDP.81. Their discovery does not depend on the brightness of the lens galaxy, so $S_B\equiv1$, as discussed in Sect.~\ref{sec:Meth}. Systems whose redshifts and velocity dispersions rely instead on a cross-match with an existing spectroscopic library are assigned to the category of that library, whatever the channel through which they were first noticed.

\section{Results and discussion}  
\label{sec:results}


\subsection{Constraints on ETG mass profiles}
Using the newly compiled lens sample, we derive the posterior probability distributions of the Jeans parameters $(\gamma,\delta,\beta)$ in three redshift intervals, following the methodology described in Section~\ref{sec:Meth}. The VDF parameters adopted for the corresponding redshift intervals are summarized in Table~\ref{Table: VDF}. At low redshift ($0<z_l<0.5$), we use the local ETG VDF from \citet{2007ApJ...658..884C}. This provides a well-calibrated low-redshift reference for the velocity-dispersion distribution of ETGs. Its use over the low-redshift interval is also motivated by the result of \citet{2011ApJ...737L..31B}, who found little evolution in the VDF between $z\sim0$ and $z\sim0.5$. For the intermediate redshift interval ($0.5<z_l<0.6$), we adopt the directly measured VDF from \citet{2017MNRAS.468...47M}, based on BOSS spectra of massive red-sequence galaxies at $z\sim0.55$. For the high-redshift interval ($0.6<z_l\lesssim1.0$), we use the LEGA-C VDF from \citet{2022ApJ...939...90T}, which provides a direct spectroscopic calibration of the VDF for quiescent galaxies at $z\sim0.6-1.0$.

\begin{table*}
\centering
\caption{Best-fit parameters of the modified Schechter function for the velocity dispersion function (VDF) adopted in this work.}
\label{Table: VDF}
\setlength{\tabcolsep}{9pt}
\footnotesize
\begin{tabular}{lccccc}
\hline\hline
Source & Redshift & $n_*$ ($\times 10^{-3}\,h^3\,\mathrm{Mpc}^{-3}$) & $\sigma_{* \rm ref}$ ($\mathrm{km~s^{-1}}$) & $\alpha$ & $\beta^{\prime}$ \\
\hline\\
\citet{2007ApJ...658..884C}  & local & $8.0 $ & $161 \pm 5$ & $2.32 \pm 0.1$ & $2.67 \pm 0.07$ \\[5pt]
\citet{2017MNRAS.468...47M}  & $0.55$ & $5.48 \pm 1.52$ & $116.40 \pm 15.12$ & $5.03 \pm 1.03$ & $2.23 \pm 0.17$ \\[5pt]
\citet{2022ApJ...939...90T}  & $0.6$--$1.0$ & $1.96^{+0.91}_{-0.37}$ & $183^{+50}_{-69}$ & $2.2^{+2.0}_{-1.2}$ & $2.8^{+1.1}_{-0.8}$ \\[5pt]
\hline
\end{tabular}
\end{table*}

\begin{figure}
  {{\includegraphics[width=3.6 in]{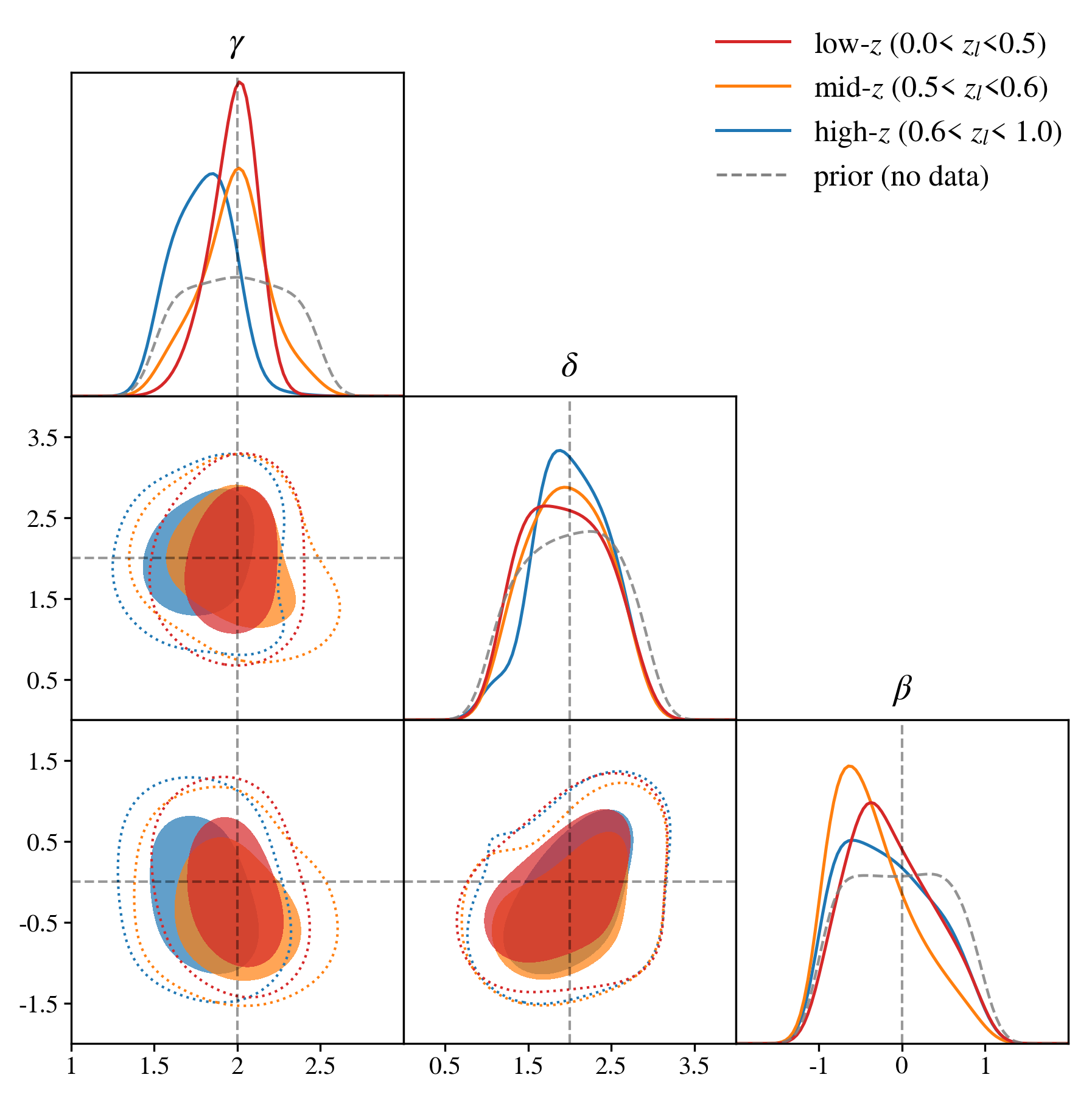}}}
  \caption{Posterior distributions of the parameters $\gamma$, $\delta$, and $\beta$ in the three redshift intervals for the fiducial VDF choice. Grey dashed lines mark the reference values $(\gamma,\delta,\beta)=(2,2,0)$. Grey dashed curves show the reference distribution obtained without any lens data. The shaded and dotted contours indicate the 68\% and 95\% credible regions, respectively. }
  \label{fig:zl_3}
\end{figure}

In the fiducial setup we adopt the category-specific magnitude windows of Sect.~\ref{sec: data_sel}. For foreground suppression we set $\varepsilon=0.2$, and place $m_{\rm sup}$ at the 10th percentile of the predicted magnitude distribution in each category. Figure~\ref{fig:zl_3} shows the resulting posteriors together with a reference distribution built without any lens data in dashed grey curves, obtained by sampling uniformly inside the prior box ($\gamma\in(1.5,2.5)$, $\delta\in(1,3)$, $\beta\in(-1,1)$) and keeping the points where $f(\gamma,\delta,\beta)$ of Eq.~(\ref{eq:f_g_d_b}) is defined. Since $f$ is defined over only part of the box, this reference distribution is not totally flat in $\delta$. Across the three redshift intervals, the $\delta$ posteriors remain broad and are centered around values close to $\delta\simeq2$, without showing a significant redshift dependence. The orbital anisotropy parameter $\beta$ is also broadly constrained and remains statistically consistent with isotropy. These parameters are therefore treated mainly as structural degrees of freedom to be marginalized over, while we focus the physical interpretation on the total density slope $\gamma$, which is the primary tracer of the effective mass structure probed by the lens. In the low-redshift bin, which contains 140 lenses, the total density slope is constrained to $\gamma = 2.00^{+0.07}_{-0.15}$, consistent with an isothermal total mass-density profile. In the intermediate-redshift bin ($0.5<z_l<0.6$), the sample contains 29 lenses and the posterior distributions are broader. We obtain $\gamma = 1.96^{+0.16}_{-0.26}$, again consistent with an isothermal profile and with the low-redshift result within the current uncertainties. For the high-redshift bin ($0.6<z_l\lesssim1.0$), which contains 31 lenses, the fiducial analysis gives $\gamma = 1.81^{+0.17}_{-0.21}$. The central value of $\gamma$ is lower than those of the low- and intermediate-redshift bins, suggesting a possible flattening of the total density profile at earlier cosmic times. The difference between the low- and high-redshift central values is $\Delta\gamma\sim 0.20$, corresponding to a significance of $\sim 1\sigma$ when the uncertainties of the two bins are combined. We therefore interpret this result as a suggestive trend rather than a definitive detection of redshift evolution.

The three intervals draw on different discovery channels and VDF calibrations, so the selection function and the VDF do not act on them in the same way. Table~\ref{tab:selection_tests} summarizes how $\gamma$ responds to each factor. Switching on the selection function leaves $\gamma$ essentially unchanged, and varying $\varepsilon$ from zero to twice its fiducial value shifts $\gamma$ by less than $0.03$ in every interval. The magnitude window matters more. In the data, lenses become rarer towards high $z_l$, and the model can explain this in two ways. Either the massive lenses needed at large distances are intrinsically rare, which the optical-depth kernel captures and which sets in earlier for larger $\gamma$, or distant lenses are too faint to pass the magnitude limit, which $S_B$ captures. A tighter window attributes more of the decline to selection and leaves less for the kernel, so the fitted $\gamma$ drops. The shift is largest at low redshift, where spectroscopically selected systems with a faint-end limit dominate. The intermediate-redshift interval is dominated by BOSS systems, whose predicted magnitudes already sit close to the targeting limit, so that the contribution from the high-$z_l$ side is removed as soon as the selection function is applied and tightening it further produces little additional shift. Finally, the $-1.0$~mag shift is included as a limiting test. It lowers $\gamma$ to $1.79$ and $1.70$ in the low- and high-redshift intervals, indicating that only a magnitude threshold this strict could substantially weaken the negative $\gamma$--$z$ trend.


\begin{table*}
\centering
\caption{Response of the total density slope $\gamma$ to the analysis settings. The first row is the fiducial analysis. The magnitude-window rows shift all magnitude thresholds, including $m_{\rm sup}$, towards brighter values by the amount indicated, and the foreground-suppression rows vary $\varepsilon$ away from its fiducial value of $0.2$. The VDF rows vary one Schechter parameter at a time within its quoted uncertainty in the interval concerned, keeping the others fixed; the last row samples $\sigma_{*\rm ref}$, $\alpha$ and $\beta^\prime$ jointly with $(\gamma,\delta,\beta)$ using split-normal priors centred on the LEGA-C measurements of \citet{2022ApJ...939...90T}. Dashes mark tests not carried out in that interval.}

\setlength{\tabcolsep}{8pt}
\footnotesize
\begin{tabular}{lccc}
\hline\hline
Treatment & $0.0<z_l<0.5$ & $0.5<z_l<0.6$ & $0.6<z_l\lesssim1.0$ \\
 & ($n=140$) & ($n=29$) & ($n=31$) \\
\hline
Fiducial & $2.00^{+0.07}_{-0.15}$ & $1.96^{+0.16}_{-0.26}$ & $1.81^{+0.17}_{-0.21}$ \\
\hline
\multicolumn{4}{l}{\textit{Selection function}} \\
No $S_B$ & $2.00^{+0.07}_{-0.15}$ & $2.00^{+0.16}_{-0.22}$ & $1.80^{+0.15}_{-0.19}$ \\
Magnitude window $-0.25$ mag & $1.97^{+0.08}_{-0.16}$ & $1.98^{+0.17}_{-0.27}$ & $1.80^{+0.19}_{-0.20}$ \\
Magnitude window $-0.5$ mag  & $1.93^{+0.12}_{-0.17}$ & $1.99^{+0.19}_{-0.28}$ & $1.77^{+0.21}_{-0.19}$ \\
Magnitude window $-1.0$ mag  & $1.79^{+0.17}_{-0.15}$ & $1.96^{+0.24}_{-0.27}$ & $1.70^{+0.23}_{-0.14}$ \\
$\varepsilon=0$   & $2.00^{+0.06}_{-0.14}$ & $1.96^{+0.16}_{-0.26}$ & $1.81^{+0.17}_{-0.20}$ \\
$\varepsilon=0.4$ & $2.00^{+0.07}_{-0.15}$ & $1.95^{+0.16}_{-0.26}$ & $1.83^{+0.16}_{-0.21}$ \\
\hline
\multicolumn{4}{l}{\textit{Velocity dispersion function}} \\
$\sigma_{*\rm ref}$ high & $2.02^{+0.07}_{-0.15}$ & $1.98^{+0.19}_{-0.29}$ & $1.95^{+0.16}_{-0.28}$ \\
$\sigma_{*\rm ref}$ low  & $1.98^{+0.07}_{-0.14}$ & $1.91^{+0.14}_{-0.23}$ & $1.63^{+0.12}_{-0.09}$ \\
$\alpha$ high(4.2)     & --- & --- & $1.84^{+0.17}_{-0.22}$ \\
$\alpha$ low(1.0)       & --- & --- & $1.80^{+0.17}_{-0.20}$ \\
$\beta^\prime$ high(3.9) & --- & --- & $1.67^{+0.14}_{-0.12}$ \\
$\beta^\prime$ low(2.0)  & --- & --- & $2.04^{+0.15}_{-0.30}$ \\
VDF marginalized & --- & --- & $1.96^{+0.22}_{-0.31}$ \\
\hline
\end{tabular}
\label{tab:selection_tests}
\end{table*}

The near-isothermal slopes recovered in the low- and intermediate-redshift bins are consistent with the long-standing result that massive ETG lenses have approximately isothermal total mass-density profiles over the radial range probed by strong lensing. This behavior is often interpreted as the outcome of the coupled evolution of baryons and dark matter: centrally concentrated stellar components, halo response to baryonic assembly, feedback-driven redistribution of gas, and subsequent merging can combine to produce an approximately scale-free total density profile even when neither the stellar nor dark-matter component is individually isothermal. In this sense, the agreement of the low- and intermediate-redshift bins with $\gamma\simeq2$ supports the view that the lensing probes a radial range where the stellar and dark-matter contributions are closely balanced in massive ETGs. The lower central value obtained in the high-redshift bin may indicate that massive ETGs at earlier epochs occupy a somewhat different structural state from their low-redshift counterparts. A shallower effective total slope at $z_l\gtrsim0.6$ would be qualitatively consistent with a picture in which the stellar mass distribution, dark-matter fraction within the lensing aperture, and halo response to baryonic processes evolve with time. Hydrodynamical simulations, however, do not yet agree on the direction of this evolution. In Horizon-AGN, AGN feedback produces shallower total slopes at earlier epochs \citep{2017MNRAS.472.2153P,2019MNRAS.483.4615P}, qualitatively consistent with our high-redshift result. In IllustrisTNG, AGN feedback makes the slopes shallower between $z\sim2$ and $z\sim1$, after which they change little \citep{2019MNRAS.490.5722W}. In Magneticum, the intrinsic slopes are steeper at higher redshift, yet the observed trend is recovered once the simulated galaxies are analysed with the observational method \citep{2017MNRAS.464.3742R}, which shows that such comparisons depend on how the effective slope is measured. In our work, given the current uncertainties, the negative $\gamma-z$ trend interpretation should be regarded as qualitative. The high-redshift bin is particularly informative because it probes an epoch where both the lens population and the VDF are less well sampled than at lower redshift. As shown in Table~\ref{Table: VDF}, the LEGA-C VDF has relatively large uncertainties. Moreover, as demonstrated by \citet{2022ApJ...939...90T}, it exhibits substantial internal shape variations across $0.6 < z_l < 1.0$. We therefore examine the sensitivity of the high-redshift result to the adopted VDF normalization and shape in the following subsection.

\subsection{Stability of the inferred density slopes}
\label{sec:Stability}

The fiducial constraints discussed above rely on the adopted VDF parameters and on the constructed lens sample. We therefore perform a set of diagnostic tests to examine how the inferred density slopes respond to a range of methodological and data-related variations. These tests are designed to separate several effects that can enter a statistical lensing analysis: the VDF calibration that sets the optical-depth kernel, the construction of the lens sample, and differences between the lens populations reached by different discovery channels.

We first examine the sensitivity of the inferred slopes to the adopted VDF calibration. In the low-redshift bin, varying $\sigma_{*\rm ref}$ within its quoted $1\sigma$ uncertainty changes $\gamma$ by only $\simeq0.04$ across the full range. The effect is larger in the intermediate-redshift bin, with changes $\Delta\gamma < 0.1$, but remains within the statistical uncertainty of the fiducial constraint. In the high-redshift bin, the broader uncertainties of the available VDF calibration lead to a larger propagated uncertainty in the inferred density slope. We therefore perform a dedicated set of high-redshift VDF tests, summarized in Table~\ref{tab:selection_tests}. These tests show that changing the low-velocity slope $\alpha$ has little effect on $\gamma_{\rm high-z}$, whereas the characteristic velocity scale $\sigma_{*\rm ref}$ and the high-velocity cutoff parameter $\beta^{\prime}$ have a more significant impact. This behavior is expected because the strong lensing optical depth is weighted toward high-velocity-dispersion galaxies, and the lensing cross section scales approximately as a high power of velocity dispersion. The high-velocity tail of the VDF is therefore a key ingredient in interpreting the high-redshift density slope. As an exploratory test of the combined VDF uncertainty, we also perform a VDF-marginalized run in which $\sigma_{*\rm ref}$, $\alpha$, and $\beta^{\prime}$ are sampled jointly with $(\gamma,\delta,\beta)$ using independent split-normal priors based on the LEGA-C measurements. This gives $\gamma_{\rm high-z}=1.96^{+0.22}_{-0.31}$, with uncertainties about 1.5 times larger than in the fixed-VDF case, and the selection function plays a visible role in this test. Without $S_B$, the same run gives $\gamma_{\rm high-z}=1.67^{+0.22}_{-0.12}$, in which case the decline of lenses towards high $z_l$ can then only be attributed to the VDF, which drives $\gamma$ towards lower values. With $S_B$, part of this decline is attributed to selection and $\gamma$ rises accordingly. This demonstrates the degeneracy between $\gamma$ and the high-velocity tail of the VDF, which dominates the systematic uncertainty of the high-redshift slope. Thus, a larger high-redshift lens sample would allow the lens analysis to be carried out in narrower redshift intervals, closer to the $\Delta z=0.1$ VDF measurements of \citet{2022ApJ...939...90T}, while maintaining sufficient statistical precision in each bin.

\begin{figure*}
\centering
  {{\includegraphics[width=5 in]{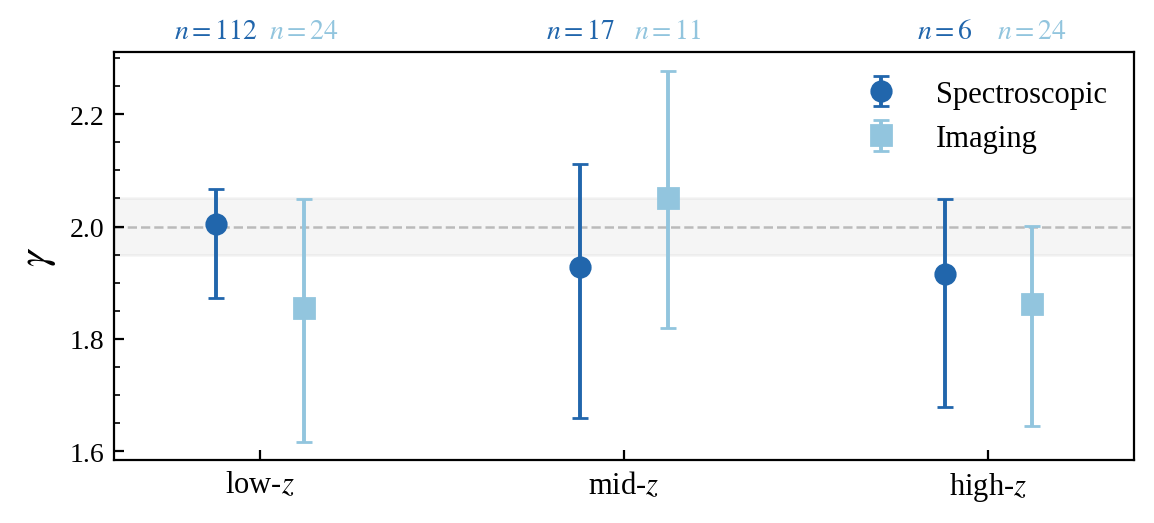}}}
  \caption{Density-slope constraints for image-selected and spectroscopically selected lenses in the three redshift intervals. The comparison tests whether the lens populations reached by the two discovery channels share a common population-level density slope. The sample sizes for each subsample are indicated at the top. The dashed line and shaded region mark the isothermal reference value $\gamma=2$. Error bars denote the 16th-84th percentile credible intervals.}
  \label{fig:discovery}
\end{figure*}


As a complementary diagnostic, we test a simple evolving-VDF prescription in the largest subsample, the low-redshift bin. We adopt
\begin{equation}
n_*(z)=n_{*,0}(1+z)^P,\qquad
\sigma_*(z)=\sigma_{*,0}(1+z)^U ,
\end{equation}
with $(P,U)=(-1,0.2)$. The density-evolution term $P=-1$ represents a mild decline in the comoving number density of ETGs with increasing redshift, consistent with the evolutionary trends reported by \citet{2007ApJ...665..265F} and \citet{2007ApJ...654..858B}. The velocity-dispersion evolution term is motivated by studies suggesting that the velocity dispersions of early-type galaxies increase with redshift, for example $\sigma \propto (1+z)^{0.19}$ at fixed stellar mass and $\sigma \propto (1+z)^{0.41}$ at fixed dynamical mass \citep{2010A&A...524A...6S}. Since other statistical studies have reported weak or insignificant evolution in $\sigma_*$, we adopt $U=0.2$ as a moderate positive evolution model rather than as a unique description of the VDF evolution. In our conditional likelihood, the absolute normalization of the VDF largely cancels between the numerator and the redshift-normalization integral. This test therefore mainly probes the impact of a mild redshift evolution in the characteristic velocity dispersion $\sigma_{* \rm ref}$. For the low-redshift bin, this evolving-$\sigma_{* \rm ref}$ prescription gives $\gamma=2.06^{+0.07}_{-0.12}$, compared with the fiducial value $\gamma=2.00^{+0.07}_{-0.15}$. The shift is therefore $\Delta\gamma\simeq0.06$, comparable to the upper $1\sigma$ uncertainty of the fiducial constraint. Although VDF evolution represents a non-negligible systematic effect for precision measurements of $\gamma(z)$, the qualitative conclusion that the low-redshift population is close to isothermal remains unchanged. For a more rigorous treatment, a full joint model of VDF and density-slope evolution should be pursued in future work.

We then examine the impact of sample construction. Repeating the analysis with only the original 160-lens sample gives $\gamma=1.99^{+0.07}_{-0.15}$, $1.94^{+0.19}_{-0.27}$, and $1.91^{+0.12}_{-0.22}$ in the low-, intermediate-, and high-redshift bins, respectively. Thus, the original sample already favors near-isothermal slopes at low and intermediate redshifts and a lower central value at high redshift. The expanded 200-lens sample mainly affects the high-redshift interval, where the number of lenses increases from 18 to 31, shifting the high-redshift central value from $\gamma\simeq1.91$ to $\gamma\simeq1.81$ and thereby excluding the isothermal $\gamma=2$ at the $1\sigma$ level. This indicates that the newly added systems increase the leverage on the high-redshift population and strengthen the apparent high-redshift flattening, while leaving the low- and intermediate-redshift results essentially unchanged. As a complementary spectroscopic quality check, we remove systems with fractional velocity-dispersion uncertainties larger than 20\%. This cut removes systems with less precise spectroscopic characterization and tests whether the inferred slopes are driven by this subset of objects. This reduces the sample from 200 to 184 lenses. The inferred slopes remain unchanged within uncertainty in the three redshift bins, indicating that the main constraints are not driven by the subset with the poorly constrained velocity dispersions.

Another important inference concerns whether lenses discovered through different strategies share a common population-level density slope. Different lens-search strategies naturally favor different regions of lens and source parameter space. Spectroscopic searches, such as those used in SLACS-like samples, preferentially identify foreground galaxies with strong absorption features and detectable background emission lines within a fixed fibre aperture, and therefore tend to select massive, relatively low-redshift lens galaxies with favorable fibre-source alignment. Image-selected searches, in contrast, are more directly sensitive to resolved arcs and multiple images, and can extend more efficiently to larger Einstein radii, higher source redshifts and lens redshifts, depending on imaging depth, angular resolution, and lens-galaxy subtraction. Our selection model accounts for the foreground-brightness dependence of each strategy, while the two strategies also reach lens galaxies of different mass and environment. Any dependence of the density slope on these properties could therefore appear as a difference in the inferred population-level $\gamma$ between the two subsamples. As shown in Figure~\ref{fig:discovery}, the image-selected and spectroscopically selected subsamples populate the redshift and space differently. In the low-redshift bin, the spectroscopically selected subsample gives $\gamma=2.01^{+0.06}_{-0.13}$, while the image-selected subsample gives $\gamma=1.85^{+0.20}_{-0.24}$. The difference in central value is within $1\sigma$, and the image-selected low-redshift subsample is substantially smaller and has a broader posterior. In the intermediate-redshift bin, both discovery channels are consistent with an isothermal slope within their larger uncertainties. In the high-redshift bin, the image-selected sample gives $\gamma=1.86^{+0.14}_{-0.22}$, while the spectroscopically selected subsample gives $\gamma=1.92^{+0.13}_{-0.24}$. The latter result is reported for completeness because the spectroscopically selected high-redshift subsample contains only six systems. The high-redshift constraint is therefore dominated in sample size by the image-selected population, but both currently available spectroscopically selected and image-selected systems indicate a similar slope in the high-redshift range. 

Taken together, the tests presented in this section demonstrate that the qualitative picture, a near-isothermal low-redshift population and a shallower high-redshift slope, is stable across a range of data-quality cuts, discovery channels, and modeling choices. The quantitative depth of the high-redshift flattening, however, remains sensitive to the assumed high-redshift VDF, particularly its high-velocity tail, and to the size and composition of the high-redshift lens sample, underscoring the need for both improved VDF measurements and larger high-redshift lens samples in future work.

\begin{table*}
\centering
\caption{Total mass-density slope constraints from previous studies and this work}
\label{Tab: sum}
\setlength{\tabcolsep}{6pt}
\footnotesize
\begin{tabular}{lccccc}
\hline\hline
Study & Sample Size & Redshift (median) & $\langle\gamma\rangle$ & Light Profile & Anisotropy \\
\hline
This work & 140 & $\sim 0.23$   & $2.00^{+0.07}_{-0.15}$   &  parametric power-law ($\delta$ free)    & $\langle\beta\rangle = -0.27^{+0.64}_{-0.43}$ \\
This work & 29  & $\sim 0.55$ & $1.96^{+0.16}_{-0.26}$   &   parametric power-law ($\delta$ free)   & $\langle\beta\rangle = -0.32^{+0.62}_{-0.42}$ \\
This work & 31  & $\sim 0.73$ & $1.81^{+0.17}_{-0.21}$   &  parametric power-law ($\delta$ free)  & $\langle\beta\rangle = -0.16^{+0.70}_{-0.58}$ \\
\hline
\multicolumn{6}{l}{Combined lensing and dynamics studies} \\
\cite{2024MNRAS.530.1474T}$^1$ & 24 & $\sim 0.5$ & $2.08 \pm 0.06$ & double S\'ersic profile     & $\langle\beta\rangle = 0.53^{+0.27}_{-0.45}$ \\
\cite{2024MNRAS.530.1474T}$^2$ & 34 & $\sim 0.2$ & $2.04 \pm 0.04$ & double S\'ersic profile     & $\langle\beta\rangle = 0.53^{+0.27}_{-0.45}$ \\
\cite{2025MNRAS.541....1S} & 56 & $\sim 0.35$ & $2.24 \pm 0.14$ & single S\'ersic profile     & $\langle\beta\rangle = -0.20^{+0.21}_{-0.19}$ \\
\cite{2016MNRAS.461.2192C}$^1$      & 118 & $\sim 0.29$ & $2.035 \pm 0.013$ & parametric power-law ($\delta$ free)  & $\beta$ prior: $0.18 \pm 0.13$ \\
\cite{2016MNRAS.461.2192C}$^2$  & 118 & $\sim 0.29$ & $2.132 \pm 0.055$ & $\gamma = \delta$          & $\beta = 0$ \\
\cite{2024ApJ...970...86S} & 7 & $\sim 0.57$ & $1.91 \pm 0.08$           & double S\'ersic profile      & $\beta = 0$ \\
\cite{2010ApJ...724..511A} & 73 & $\sim 0.2$ & $2.078 \pm 0.027$ & de Vaucouleurs profile       & $\beta = 0$ \\
\cite{2009ApJ...703L..51K} & 58 & $\sim 0.2$ & $2.085^{+0.025}_{-0.018}$ & Hernquist & $\beta = 0$ \\
\cite{2006ApJ...649..599K} & 15 & $\sim 0.2$ & $2.01^{+0.02}_{-0.03}$ & Hernquist                   & $\beta = 0$ \\
\cite{Sonnenfeld_2024} & 59 & $\sim 0.19$ & $1.99 \pm 0.03$    & no explicit light fit ($R_{\rm e}$-weighted) & $\beta = 0$ \\
\cite{10.1111/j.1365-2966.2011.18842.x} & 16 & $\sim 0.16$ & $2.074^{+0.043}_{-0.041}$ & non-parametric HST light    & $\beta = 0$ \\
\cite{2018MNRAS.475.2403L} & 15 & $\sim 0.16$ & $2.1 \pm 0.05$     & S\'ersic/de Vaucouleurs     & $\beta$-insensitive radius (no $\beta$ fit) \\
\cite{2023MNRAS.521.6005E}$^1$ & 123 & $\sim 0.3$ & $2.024^{+0.019}_{-0.021}$ & de Vaucouleurs profile      & $\beta$ from source L\&D research \\
\hline
\multicolumn{6}{l}{Dynamics-only studies} \\
\cite{Mozumdar2026}$^1$ & 277 & $\sim 0.044$ & $2.260 \pm 0.013$ &
MGE light model & $\sigma_\theta/\sigma_r \sim \mathcal{U}\!\left[\mathcal{R}(q),\,2\right]$ \\
\cite{Mozumdar2026}$^2$ & 163 & $\sim 0.44$ & $2.190 \pm 0.013$ &
MGE light model & $\sigma_\theta/\sigma_r \sim \mathcal{N}(1.0,0.07)$ \\
\cite{2023MNRAS.522.3602D}$^1$ & 64  & $\sim 0.35$ & $2.01 \pm 0.04$          & MGE light model                         & $\beta$ free (constant per galaxy) \\
\cite{2023MNRAS.522.3602D} $^2$& 28  & $\sim 0.3$  & $2.23 \pm 0.04$          & MGE light model                         & $\beta$ free (constant per galaxy) \\
\cite{2023MNRAS.522.3602D}$^3$ & 150 & $\sim 0$    & $2.25 \pm 0.02$          & MGE light model                         & $\beta$ free (constant per galaxy) \\
\cite{2018MNRAS.476.4543B} & 21 & $\sim 0$    & $2.24 \pm 0.05$          & MGE light model                         & $\beta$ free (constant per galaxy) \\
\cite{2017MNRAS.467.1397P}     & 258 & $\sim 0$    & $2.193 \pm 0.016$        & MGE light + (M/L)$_*$ from pPXF         & $\beta$ free (constant per galaxy) \\
\hline
\multicolumn{6}{l}{Lensing-only studies} \\
\cite{2024ApJ...970...86S}$^2$              & 7   & $\sim 0.57$ & $1.95 \pm 0.09$          & N/A                                      & N/A \\
\cite{2023MNRAS.521.6005E}$^2$   & 57  & $\sim 0.3$  & $2.095^{+0.030}_{-0.029}$ & N/A                                     & N/A \\
\hline
\end{tabular}
\tablefoot{For \cite{2024MNRAS.530.1474T}, the quoted anisotropy $\langle\beta\rangle$ is derived from a single joint fit to the full SLACS+SL2S+BELLS sample of 77 lenses, and is applied to both subsamples listed. For the two \cite{Mozumdar2026} entries, the ratio $\sigma_\theta/\sigma_r$ is assumed to be constant within each galaxy and is fitted independently for each galaxy; the quoted
$\mathcal{U}[\mathcal{R}(q),2]$ and $\mathcal{N}(1.0,0.07)$ are the priors on this constant. Here $q$ is the intrinsic axial ratio and
$\mathcal{R}(q)=\sqrt{0.3+0.7q}$. Superscript labels (e.g., $^{1}$, $^{2}$) denote distinct analysis configurations reported within the same work, for example measurements obtained from different subsamples or from alternative modeling choices.}
\end{table*}

\begin{figure*}
\centering
 {{\includegraphics[width=6.8 in]{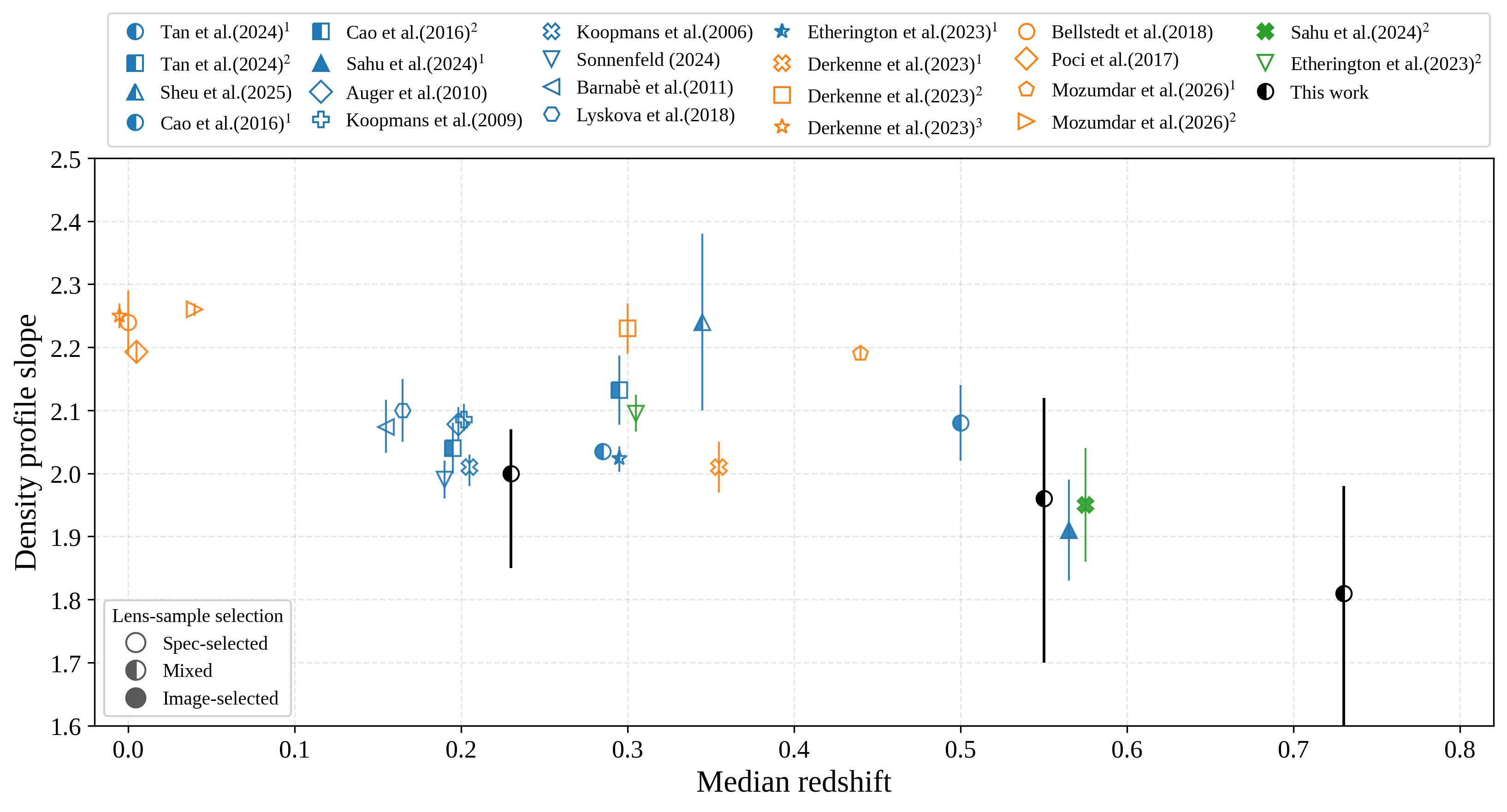}}}
\caption{Comparison of total mass-density slopes $\langle\gamma\rangle$ for different samples. Blue, orange, and green symbols represent joint lensing-dynamics, dynamics-only, and lensing-only studies, respectively. Black points show the three redshift subsamples from this work. Error bars indicate $1\sigma$ uncertainties.
To highlight differences in the underlying lens samples, the marker fill style encodes the lens-selection method: open symbols denote studies based on spectroscopically selected lenses (e.g.\ SLACS, BELLS, BELLS~GALLERY), filled symbols denote studies based on image-selected lenses (e.g.\ SL2S, AGEL), and half-filled symbols denote studies that combine both selection types within a single sample (e.g.\ SLACS+SL2S+BELLS). Dynamics-only studies do not rely on imaging and are plotted with open symbols by default. Note that several literature measurements share nearly identical median redshifts, a small horizontal jitter of $\pm0.005$ in redshift was applied to avoid overlapping.
}
\label{fig:sum}
\end{figure*}

\subsection{Implications for ETG density-profile evolution}
\label{sec:implications}

The redshift evolution of the total mass-density slope of massive early-type galaxies has been investigated using a variety of lensing, dynamical, and joint lensing-dynamics approaches. Dynamical studies of nearby ETGs often infer slightly steeper inner total slopes, with typical values around $\gamma\simeq2.2$ \citep{2015ApJ...804L..21C,2017MNRAS.467.1397P,2019MNRAS.490.2124L}, whereas strong lensing and joint lensing-dynamics studies of massive lenses commonly find slopes close to isothermal over the radial range probed by the Einstein radius. The reported normalization and possible redshift dependence of the slope, however, vary among different studies. This diversity is expected because different analyses probe different samples, radial ranges, selection functions, and modelling assumptions. Table~\ref{Tab: sum} and Figure~\ref{fig:sum} place our fiducial constraints in the context of previous measurements. The low- and intermediate-redshift results from this work, $\gamma=2.00^{+0.07}_{-0.15}$ and $\gamma=1.96^{+0.16}_{-0.26}$, are consistent with the slopes reported at similar redshifts by many studies of massive ETG lenses. The high-redshift result from this work, $\gamma=1.81^{+0.17}_{-0.21}$, lies on the shallower side of the literature distribution, extending the statistical comparison into a redshift range where existing lens constraints remain sparse. The direction of this negative trend in $\gamma$ with redshift is consistent with a recent dynamics-only analysis \citep{Mozumdar2026}, which used spatially resolved MUSE-DEEP stellar kinematics together with a local MaNGA comparison sample and reported $\langle\gamma\rangle=2.26\pm0.01$ at $\langle z\rangle=0.04$ and $\langle\gamma\rangle=2.19\pm0.01$ at $\langle z\rangle=0.44$, corresponding to ${\rm d}\gamma/{\rm d}z\simeq -0.20\pm0.03$. The higher absolute normalization of $\gamma$ relative to the slopes inferred here and in many lensing studies likely reflects differences in galaxy samples, radial ranges, and modelling assumptions. At $z\lesssim0.55$, \citet{2021MNRAS.506.3691D} found no significant evolution of the dynamically inferred slope ($\gamma=2.11\pm0.03$ for 90 ETGs at $0.29<z<0.55$), consistent with the near-constant slopes of our low- and intermediate-redshift bins; the decline in our results appears mainly in the $0.6<z_l\lesssim1.0$ interval, beyond the redshift range covered by current dynamical samples. Together, these comparisons highlight the importance of method-dependent slope definitions and radial sensitivity.

The scatter among literature measurements also reflects differences in the radial range probed and the observables used by each method, as summarized in Table~\ref{Tab: sum}. Our analysis connects the Einstein radius to a reference stellar velocity dispersion through a Jeans-based mapping, with the kinematic information entering statistically through the adopted VDF. Unlike studies that fit a slope to each lens, however, we infer a single population-level slope from the distribution of lens redshifts conditioned on the observed Einstein radii and source redshifts. Lensing-only analyses constrain the projected mass distribution within the Einstein radius, but the inferred density slope depends on the adopted mass model and priors and is subject to the mass-sheet degeneracy. Dynamics-only studies, in contrast, measure luminosity-weighted velocity moments of normal galaxies, typically out to the effective radius or beyond, so that the inferred slopes depend on the stellar orbital structure, the adopted light model, and the radial range covered by the kinematic data. Such samples are free of lens-finding biases, but are selected through the surface-brightness and signal-to-noise requirements of spatially resolved spectroscopy. Their higher values of $\gamma$ may partly reflect differences in radial coverage and slope definition. Comparisons of redshift trends also require accounting for changes in sample properties, selection, and spatial resolution with redshift. Joint lensing-dynamics analyses combine these complementary constraints and partly break their respective degeneracies. They provide a useful comparison for our results, although differences in data quality, spatial coverage, and the treatment of the stellar light distribution, aperture corrections, orbital anisotropy, and symmetry assumptions should be considered when comparing the inferred slopes.

In our framework, the stellar luminosity density is described by an effective power-law slope, $\nu(r)\propto r^{-\delta}$. This parametrization is commonly used in analytical Jeans-based lensing calculations \citep[e.g.,][]{2006ApJ...649..599K,2010ApJ...708..750S,2016MNRAS.461.2192C,2025A&A...694A.196G}. While individual lens light profiles are observable, a single power law yields only one effective value: for Sérsic profiles with $2\lesssim n\lesssim6$, the logarithmic slope of the deprojected luminosity density increases from $\sim 2.0$--$2.6$ at $0.5R_{\rm e}$ to $\sim 2.6$--$2.8$ at $R_{\rm e}$. We therefore treat $\delta$ as a nuisance parameter of the Jeans mapping and marginalize over $1\leq\delta\leq3$, which covers this range. Orbital anisotropy is another important source of methodological variation among published studies. Many joint lensing-dynamics analyses adopt simplified treatments of the anisotropy, often fixing $\beta=0$ or imposing an informative prior. In our fiducial analysis, $\beta$ is allowed to vary and is marginalized over together with $\gamma$ and $\delta$. The resulting $\beta$ posteriors are broad and statistically consistent with isotropy in all three redshift intervals. As a check against the commonly adopted isotropic assumption, we also repeated the analysis with $\beta=0$. This changes the inferred density slopes by only $\Delta\gamma \lesssim 0.05$ relative to the fiducial case in all redshift bins and does not alter the redshift ordering of the results. Thus, within the current statistical precision, the qualitative comparison with previous studies is not driven by whether the orbital anisotropy is fixed to zero or marginalized over. The importance of anisotropy and symmetry assumptions has also been emphasized in more detailed dynamical studies. For example, \citet{10.1111/j.1365-2966.2011.18842.x} used axisymmetric, two-integral dynamical models of 16 SLACS lenses and obtained density-slope estimates systematically lower by $\sim0.05$ than those inferred by \citet{2010ApJ...724..511A} using a spherical and isotropic Jeans model. As discussed by \citet{2025A&A...697A..95S}, this difference is comparable to or larger than the statistical uncertainty on the mean SLACS slope, illustrating the relevance of dynamical modelling assumptions. More recently, \citet{2024MNRAS.530.1474T} compared different anisotropy prescriptions, including spatially constant and Osipkov--Merritt forms, and showed that the inferred consistency with a single power-law total profile can depend on the assumed $\beta(r)$ model. Simulations further suggest that $\beta$ may evolve with galaxy mass and redshift, with more tangentially biased orbits at higher redshift and more radially biased orbits at lower redshift \citep{2017MNRAS.469.1824X}. Our $\beta$ posteriors are too broad to test this directly, and future spatially resolved kinematics will be needed to constrain anisotropy and density-slope evolution simultaneously.


Physically, the near-isothermal slopes found at low and intermediate redshift are commonly interpreted as a manifestation of the bulge-halo conspiracy. The stellar component of massive ETGs is centrally concentrated, whereas the dark-matter halo is intrinsically shallower over the same radial range. Their combined contribution can therefore produce a nearly power-law total density profile with $\gamma\simeq2$ around the Einstein radius, even though neither component is individually isothermal. This balance can be shaped by baryonic contraction, stellar mass assembly, feedback-regulated gas removal, and subsequent merging. The lower central value of $\gamma$ in the high-redshift bin may point to a different balance between stellar and dark-matter contributions at earlier epochs. A shallower effective total slope can arise if the stellar component is less centrally dominant within the lensing aperture, if the dark-matter fraction within the Einstein radius is higher, or if the halo response to baryonic processes differs from that at later times. In galaxy-formation models, the outcome of these processes depends strongly on how stellar and AGN feedback are implemented, and strong-lens slopes are sensitive to this choice. \citet{2021MNRAS.504.3455M} showed that only some of the feedback models in the EAGLE simulations reproduce the distribution of total slopes at the Einstein radius observed for SLACS and BELLS lenses. The redshift evolution of the slope offers a further test. Several cosmological hydrodynamical simulations yield nearly isothermal slopes for massive ETGs at low redshift, but differ in how these slopes evolve towards $z\sim1$. In Horizon-AGN, the total density slope steepens from $z=2$ to $z=0$, and AGN feedback is needed to bring the slopes close to the observed values \citep{2017MNRAS.472.2153P,2019MNRAS.483.4615P}. In IllustrisTNG, the ETG population selected at each redshift shows nearly constant slopes below $z\approx1$ with a mild steepening towards $z=0$ \citep{2019MNRAS.490.5722W}. In Magneticum and Illustris, by contrast, the slopes are steeper at higher redshift and become shallower towards $z=0$ \citep{2017MNRAS.464.3742R,2017MNRAS.469.1824X}, which \citet{2017MNRAS.464.3742R} attribute to the dominance of gas-poor mergers at late times. Since these predictions diverge increasingly with redshift, our high-redshift bin probes the range in which the models are best distinguished. The shallower slope obtained at $0.6<z<1.0$ follows the direction predicted by Horizon-AGN and has the same sign as the redshift dependence found for SL2S lenses \citep{2013ApJ...777...98S}. Such a steepening with cosmic time supports a modest contribution from dissipative processes, since purely dry mergers make the density profile shallower \citep{2014ApJ...786...89S}. Improved calibration of the high-velocity tail of the VDF at $z\gtrsim0.6$ and larger high-redshift lens samples will be needed to determine the amplitude of this evolution and to distinguish among these formation scenarios.

\section{Conclusion}
\label{sec:Conclusion}

In this work, we compiled a sample of 200 galaxy-scale strong lensing systems and applied a statistical strong lensing framework to investigate the effective mass structure of early-type galaxy (ETG) lenses. By combining the lens-redshift distribution with VDF calibrations and a spherical Jeans-based mapping between the Einstein radius and the velocity-dispersion scale, we constrained the total mass-density slope $\gamma$ in three redshift intervals, marginalizing over the luminosity-density slope $\delta$ and the orbital anisotropy parameter $\beta$. Under the fiducial VDF assumptions, the low- and intermediate-redshift bins yield total density slopes close to the isothermal value, with $\gamma=2.00^{+0.07}_{-0.15}$ at $0<z_l<0.5$ and $\gamma=1.96^{+0.16}_{-0.26}$ at $0.5<z_l<0.6$. For the high-redshift bin, $0.6<z_l\lesssim1.0$, we obtain $\gamma=1.81^{+0.17}_{-0.21}$, which excludes the isothermal value at the $1\sigma$ level. The lower high-redshift value suggests a flattening of the effective total density profile at earlier cosmic epochs. This trend is strengthened by the expanded sample: relative to the original 160-lens sample, the number of lenses at $0.6<z_l\lesssim1.0$ increases from 18 to 31, and the high-redshift slope deviates from isothermal.

Our sample is compiled from surveys with different selection criteria. We therefore incorporated a selection function into the lens-redshift likelihood, modelled separately for each system class (Sect.~\ref{sec: data_sel}). As shown in Eq.~(\ref{eq:like_SB}), our likelihood is conditioned on the observed Einstein radius and source redshift, so the inference depends only on $S_B(z_l)$. Including $S_B$ has little effect on the fiducial slopes, and the most influential factor is the magnitude window: tightening the thresholds affects $\gamma$ differently in the three redshift intervals, and only the $-1.0$~mag stress test substantially weakens the redshift trend (Table~\ref{tab:selection_tests}). The role of $S_B$ is most evident when the VDF parameters are marginalized over: it then raises the high-redshift slope from $\gamma\simeq1.67^{+0.22}_{-0.12}$ to $1.96^{+0.22}_{-0.31}$, since part of the decline in lens numbers towards high $z_l$ is attributed to selection rather than to the VDF.

The inferred slopes remain stable in most of our diagnostic tests (Sect.~\ref{sec:Stability}). Fixing $\beta=0$ changes the inferred density slopes by only $\Delta\gamma\lesssim0.05$ and does not alter the qualitative redshift ordering. The results are also stable against removing systems with large fractional velocity-dispersion uncertainties, and spectroscopically and image-selected lenses yield consistent slopes in all three intervals (Fig.~\ref{fig:discovery}). The dominant systematic uncertainty arises from the VDF, especially at high redshift. In particular, the high-redshift result is sensitive to the characteristic velocity dispersion and the high-velocity cutoff of the LEGA-C VDF, which shift the slope by amounts comparable to its statistical uncertainty. Thus, the present analysis identifies high-redshift VDF calibration as a key requirement for turning the trend found here into a more precise measurement of ETG density-profile evolution.

The near-isothermal slopes recovered at low and intermediate redshift are consistent with the established picture that massive ETG lenses have approximately isothermal total mass-density profiles over the radial range probed by strong lensing. The lower central value at high redshift follows the direction of the evolution found for SL2S lenses \citep{2013ApJ...777...98S}, the dynamics-only results based on MUSE-DEEP and MaNGA ETGs \citep{Mozumdar2026}, and the prediction of Horizon-AGN \citep{2019MNRAS.483.4615P}. In a two-component picture, a shallower effective total slope could reflect a smaller relative stellar contribution within the Einstein radius at earlier epochs. Subsequent central mass growth would then move the profile toward the near-isothermal structure seen at lower redshift. Gas-poor mergers and dissipative growth can affect this balance differently: the former tend to add stars at larger radii, whereas the latter can increase the central stellar concentration. A modest contribution from dissipative growth below $z\sim1$ is therefore compatible with the trend found here (Sect.~\ref{sec:implications}).


The number of known galaxy-scale strong lenses is now increasing rapidly. The Euclid Quick Data Release alone yielded about 500 lens candidates over 63~deg$^2$, and the first Euclid data release is expected to provide several thousand high-confidence candidates \citep{2026A&A...711A..26E}. Ground-based searches in the DESI Legacy Imaging Surveys and UNIONS have identified large numbers of further candidates \citep{DESI_LS_lenses,2026ApJ...999....1S}, and Euclid, the Rubin LSST, which began its ten-year survey in 2026, and the recently launched Roman Space Telescope are together expected to increase the number of known lenses to $\gtrsim10^5$ \citep{2015ApJ...811...20C,Roman_forecast}. Spectroscopic programmes such as DESI and 4MOST will supply lens and source redshifts for large numbers of these systems \citep{DESI_SLF_II,2023Msngr.190...49C}. In addition, CSST is expected to detect approximately $10^5$ galaxy-galaxy strong lensing systems in simulations \citep{2024MNRAS.533.1960C}. Its optical multi-band imaging will be complementary to Euclid, LSST, and Roman for lens identification and photometric characterization, extending strong-lens studies toward broader ranges of lens mass, image separation, and redshift. Compared with ground-based surveys, the space-based angular resolution of CSST is particularly valuable for resolving smaller-separation galaxy-galaxy lenses, while its $ugriz$ photometry can improve the characterization of both lens and source galaxies. CSST slitless spectroscopy can provide useful redshift and emission-line information over a wide field, while precise stellar velocity dispersions for detailed dynamical analyses will require higher-resolution spectroscopy or cross-matches with surveys such as SDSS, DESI, and future spectroscopic facilities. The simulated CSST strong lensing catalogue of \citet{2024MNRAS.533.1960C} has already been used in several follow-up studies, demonstrating its value as a testbed for future lens searches and modeling pipelines. For example, \citet{Li_2024} applied the mock CSST images to train a machine-learning lens finder, while \citet{2025MNRAS.540.3121C} used them to benchmark a fast lens-modeling algorithm for large survey samples. In the context of the present work, we focus on the high-quality ''golden'' subset with signal-to-noise ratios exceeding 100 in stacked observations, corresponding to roughly $4,000$ systems in the CSST mock catalogue \citep{2024MNRAS.533.1960C}. A simplified forecast based on this subset suggests that the typical statistical uncertainty in $\gamma$ could be reduced by a factor of order $\sim5$ relative to current samples. More importantly, such samples would enable narrower redshift binning, improved tests of discovery-channel dependence, and better calibration of the high-redshift VDF. 

The statistical approach developed in our paper is suited to future samples. Applying the method to new surveys will require a characterization of their lens-redshift-dependent selection, including the magnitude limits and other foreground-dependent effects relevant to each discovery and follow-up strategy. Lenses from different programmes can then be analysed within the same framework. This approach provides a promising route toward population-level measurements of ETG mass-structure evolution across cosmic time.

\begin{acknowledgements}
We thank the anonymous referee for constructive comments that improved this paper. This work is supported by Beijing Natural Science Foundation No.~1242021; the National Natural Science Foundation of China (Nos.~12021003, 12203009, 12433001); and the Strategic Priority Research Program of the Chinese Academy of Sciences, Grant No.~XDB23000000. 

We acknowledge the use of \textsc{PyAutoLens} \citep{2021JOSS....6.2825N, 2024MNRAS.532.2441H}, \texttt{emcee} \citep{2013PASP..125..306F}, and Astropy \citep{2013A&A...558A..33A,2018AJ....156..123A,2022ApJ...935..167A}.

This research used data from DESI Data Release~1 \citep{DESI_DR1} and its \texttt{FastSpecFit} value-added catalogue \citep{FastSpecFit}, and images from the third public data release of the HSC Subaru Strategic Program \citep{2022PASJ...74..247A}. This work also made use of data from SDSS and SDSS-III.
\end{acknowledgements}

\bibliographystyle{aa}
\bibliography{ref}

\begin{appendix}
\begingroup

\section{Calculation of the selection function $S_B$}
\label{app:SB}

This appendix describes how the selection function $S_B$ of Eq.~(\ref{eq:SB}) is evaluated. For each system and each trial lens redshift $z_l$, the reference velocity dispersion $\sigma_{\rm ref}$ required to reproduce the observed Einstein radius (Eq.~\ref{eq:R}) sets the expected absolute magnitude of the lens galaxy through the Faber-Jackson relation. Once distance, the redshifting of the spectrum, and stellar population evolution are taken into account, this absolute magnitude gives the mean total apparent magnitude $m_{\rm lens}$ of Eq.~(\ref{eq:mlens}), from which both factors of $S_B$ are evaluated. The calibration of the absolute magnitude is described in Appendix~\ref{app:sigma_to_magnitude}, and the computation of the distance modulus $\mu$, the $K$-correction, and the evolution correction $E$ in Appendix~\ref{app:absolute_to_apparent}. Appendix~\ref{app:SB_win} treats the magnitude windows and the foreground suppression. The tabulated corrections are available at \url{https://github.com/Jiang-Xin-yue/lens-statistics}.

\subsection{Absolute magnitude at fixed velocity dispersion}
\label{app:sigma_to_magnitude}

The absolute magnitude $M_{\rm lens}$ in Eq.~(\ref{eq:mlens}) is the mean absolute magnitude at fixed velocity dispersion. We calibrate it with the analysis of \citet{Bernardi_PaperII}, who modelled the joint distribution of absolute magnitude, effective radius, and velocity dispersion of about 9000 SDSS early-type galaxies as a multivariate Gaussian. Only the part of this distribution that relates $M$ to $V=\log_{10}[\sigma/(\mathrm{km\,s^{-1}})]$ enters our calculation. Conditioning the Gaussian on $V$ gives a mean absolute magnitude
\begin{equation}
\langle M\mid V\rangle = M_* + A\,(V-V_*), \qquad A=\rho_{VM}\,\frac{\sigma_M}{\sigma_V},
\label{eq:fj_conditional_mean}
\end{equation}
with an intrinsic scatter
\begin{equation}
w=\sigma_M\sqrt{1-\rho_{VM}^{2}},
\label{eq:fj_conditional_scatter}
\end{equation}
where $M_*$ and $V_*$ are the mean values of the calibrating distribution, $\sigma_M$ and $\sigma_V$ are the marginal dispersions in $M$ and $V$, and $\rho_{VM}$ is their correlation coefficient. The slope $A$ thus describes how the mean absolute magnitude changes with velocity dispersion. The scatter $w$ quantifies the spread in luminosity that remains among galaxies of the same velocity dispersion.

Table~\ref{tab:bernardi} lists the calibration parameters adopted for different bands, together with the resulting \(A\) and \(w\). The SDSS magnitude window is calibrated with the \(r\)-band relation, the BOSS LOWZ and CMASS windows with the \(r\)- and \(i\)-band relations, respectively, and the DESI window with the \(z\)-band relation. The conditional relation above therefore supplies the mean absolute magnitude and scatter required for the apparent-magnitude prediction.

\begin{table}
\centering
\caption{Parameters of the conditional relation between velocity dispersion and absolute magnitude. $M_*$, $V_*$, $\sigma_M$, $\sigma_V$, and $\rho_{VM}$ are taken from Table~1 of \citet{Bernardi_PaperII}, and $A$ and $w$ are computed from them using Eqs.~(\ref{eq:fj_conditional_mean}) and (\ref{eq:fj_conditional_scatter}).}
\label{tab:bernardi}
\setlength{\tabcolsep}{4pt}
\footnotesize
\begin{tabular}{lccccccc}
\hline\hline
Band & $M_*$ & $V_*$ & $\sigma_M$ & $\sigma_V$ & $\rho_{VM}$ & $A$ & $w$ \\
     & (mag) & (dex) & (mag) & (dex) & & (mag\,dex$^{-1}$) & (mag) \\
\hline
$r$ & $-21.15$ & 2.200 & 0.841 & 0.111 & $-0.774$ & $-5.864$ & 0.533 \\
$i$ & $-21.49$ & 2.201 & 0.851 & 0.110 & $-0.781$ & $-6.042$ & 0.531 \\
$z$ & $-21.83$ & 2.200 & 0.845 & 0.110 & $-0.782$ & $-6.007$ & 0.527 \\
\hline
\end{tabular}
\end{table}

\subsection{Distance modulus, $K$-correction, and luminosity evolution}
\label{app:absolute_to_apparent}

$M_*$ obtained from Appendix~\ref{app:sigma_to_magnitude} refers to a calibration at zero redshift, and we correct the predicted absolute magnitude for distance, redshifting, and evolution at the lens redshift below. The distance modulus is $\mu(z)=5\log_{10}[D_L(z)/10\,\mathrm{pc}]$, where $D_L$ is the luminosity distance in the flat $\Lambda$CDM cosmology adopted in Sect.~\ref{sec:Meth}. The $K$-correction and the evolution correction $E$ are calculated from the same age-dependent model spectra. We use the simple stellar population (SSP) models of \citet{BC03}, which combine stellar spectra along evolutionary tracks to predict the integrated light of a population formed in a single burst. The stellar spectra are supplied by the STELIB library \citep{STELIB}, while a Chabrier initial mass function \citep{Chabrier_IMF} sets the relative numbers of stars formed at different masses. We adopt solar metallicity and assume a single burst of star formation at $z_f=2$ \citep[e.g.][]{vanDokkum_vanderMarel2007,Bernardi_PaperIII}. The age of the stellar population observed at redshift $z$ is then $t(z)=t_U(z)-t_U(z_f)$, where $t_U(z)$ is the age of the Universe at redshift $z$.

For AB magnitudes and a photon-counting detector, the same-band $K$-correction is written as \citep{Hogg_KCorrection}
\begin{equation}
K(z)=-2.5\log_{10}\!\left[
\frac{1}{1+z}\,
\frac{\int \lambda\,\ell_\lambda[\lambda/(1+z);t(z)]\,T(\lambda)\,\mathrm{d}\lambda}
{\int \lambda\,\ell_\lambda[\lambda;t(z)]\,T(\lambda)\,\mathrm{d}\lambda}
\right],
\label{eq:K_correction}
\end{equation}
where $\ell_\lambda(\lambda;t)$ is the rest-frame spectral luminosity density at age $t$. $T(\lambda)$ is the total response of filter, optics, and detector, for which we adopt the SDSS $r$- and $i$-band response functions \citep{SDSS_FilterResponse} and the DECam $z$-band bandpass used in DES Y3 \citep{DECam_DESY3_Response}. The numerator and denominator give the signal through the same filter from the redshifted and the rest-frame spectrum at age $t(z)$, respectively.

Since $M_*$ in Table~\ref{tab:bernardi} refers to zero redshift, the evolution correction compares the luminosity in the same rest-frame band at ages $t(z)$ and $t(0)$,
\begin{equation}
E(z)=-2.5\log_{10}\!\left\{\frac{L_T[t(z)]}{L_T[t(0)]}\right\}, \,
L_T(t)=\int \lambda\,\ell_\lambda(\lambda;t)\,T(\lambda)\,\mathrm{d}\lambda ,
\label{eq:E_correction}
\end{equation}
where $L_T$ is the band-weighted luminosity at age $t$. Stellar populations fade as they evolve passively, so lens galaxies at higher redshift are intrinsically brighter than at $z=0$, which gives $E(z)<0$.

The normalization integral in Eq.~(\ref{eq:like_S}) extends to $z_s$, beyond the highest lens redshift in our sample ($z_l=1.004$). Above $z=1.05$, where the population age approaches zero towards $z_f$, we hold the age fixed and continue to evaluate $\mu$ and $K$ at the actual redshift. Tabulated values of $\mu$, $K$, and $E$ are provided in the repository.

\subsection{Survey magnitude limits and foreground suppression}
\label{app:SB_win}

Since the predicted $m_{\rm lens}$ is a total magnitude, the target-selection limits of each parent spectroscopic sample must be expressed in terms of total magnitudes before the two can be compared. We describe how the survey-specific magnitude definitions are treated for SDSS, BOSS, and DESI in turn.

SDSS defines its limits in Petrosian magnitudes, which recover about 80\% of the total flux of a de Vaucouleurs profile \citep{2002AJ....124.1810S}, so that
\begin{equation}
m_{\rm Petro}-m_{\rm total}=-2.5\log_{10}(0.8)\simeq0.24 .
\label{eq:petro_offset}
\end{equation}
Shifting both limits of the adopted Petrosian window, $14.5<r<19.5$, by $-0.24$~mag therefore gives a total-magnitude window of $14.26<r<19.26$. We apply this window to all SDSS systems, whose assignment to the parent sample is described in Sect.~\ref{sec: data_sel}.\footnote{The adopted bright limit of 14.5 approximates the bright-end exclusion that SDSS applies on fibre magnitudes \citep{2002AJ....124.1810S}.}

BOSS, in contrast, selects its targets on cmodel magnitudes. These estimate the total model flux from a combined fit of de Vaucouleurs and exponential profiles \citep{SDSS_cmodel} and can therefore be used directly as total magnitudes. Since the BOSS galaxy sample comprises both the LOWZ and the CMASS selections, we adopt the LOWZ window $16.0<r<19.6$ and the CMASS window $17.5<i<19.9$ \citep{BOSS_TargetSelection}. Each window is compared with the predicted magnitude distribution in its band, and the larger of the two resulting probabilities is taken as the window factor for \texttt{SPEC\_BOSS} systems.

The DESI threshold, $z_{\rm fibre}<21.6$ \citep{DESI_LRG_TargetSelection}, instead applies to the flux within the fibre. Converting it into a total-magnitude limit requires the fraction $\eta$ of the galaxy light that enters the fibre, which depends on the angular size of the lens galaxy and hence on $z_l$ and $\sigma_{\rm ref}$. We obtain this size from the size-velocity dispersion relation of \citet{Bernardi_PaperII} and compute $\eta$ as the fraction of a de Vaucouleurs profile of each lens, convolved with a Gaussian of $1''$ FWHM, that falls within the $0.75''$ fibre radius, following the fibre-flux convention of the Legacy Surveys\footnote{\url{https://www.legacysurvey.org/dr9/catalogs/}}. Since the fibre collects only part of the light, the DESI threshold corresponds to an effective total-magnitude limit
\begin{equation}
m_{{\rm lim},z}=21.6+2.5\log_{10}\eta ,
\label{eq:mlim_desi}
\end{equation}
which we apply to the \texttt{SPEC\_DESI} category without a bright-end cut. The size-relation parameters and the aperture-correction table are provided in the repository.

For \texttt{IMAGING\_SPEC\_FOLLOWUP} systems, which were observed with dedicated spectroscopic follow-up rather than drawn from a magnitude-limited sample, the window factor is set to unity, and for \texttt{SOURCE\_SPEC\_FOLLOWUP} systems $S_B\equiv1$ (Sect.~\ref{sec: data_sel}). The foreground-suppression factor of Eq.~(\ref{eq:SB}) is evaluated with the predicted total $i$-band magnitude of the lens galaxy, the $i$-band scatter $w$ from Table~\ref{tab:bernardi}, and the fiducial amplitude $\varepsilon=0.2$. We set $m_{\rm sup}$ to the 10th percentile of the $i$-band magnitudes predicted from the measured velocity dispersions and redshifts. These percentiles are computed once from the full sample, separately for subsets of the categories in Table~\ref{tab:selection_categories} defined by targeting programme and discovery route, and the adopted values are provided in the repository. The product of the magnitude window and the foreground suppression is the selection function $S_B$ that enters the conditional likelihood of the main text.

\endgroup
\end{appendix}


\end{document}